\documentclass[]{aa}
\usepackage{color }
\usepackage{natbib}
\usepackage{amsmath}
\usepackage{amssymb}
\usepackage{graphicx}
\graphicspath{{pictures/}}
\DeclareGraphicsExtensions{.png}
\newcommand{\bl}[1]{\mbox{\boldmath$ #1 $}}

\begin{document}
\title {Evolution of dust in protoplanetary disks of eruptive stars}

\author{Eduard I. Vorobyov\inst{1,2}, Aleksandr M. Skliarevskii\inst{2}, Tamara Molyarova\inst{3}, Vitaly Akimkin\inst{3}, Yaroslav Pavlyuchenkov\inst{3}, \'Agnes K\'osp\'al\inst{4,5,6}, Hauyu Baobab Liu\inst{7}, Michihiro Takami\inst{7}, Anastasiia Topchieva\inst{3}}

\institute{ 
University of Vienna, Department of Astrophysics, Vienna, 1180, Austria \\
\email{eduard.vorobiev@univie.ac.at} 
\and
Research Institute of Physics, Southern Federal University, Roston-on-Don, 344090 Russia 
\and 
Institute of Astronomy, Russian Academy of Sciences, 48 Pyatnitskaya St., Moscow, 119017, Russia
\and
Konkoly Observatory, Research Centre for Astronomy and Earth Sciences, E\"otv\"os Lor\'and Research Network (ELKH), Konkoly-Thege Mikl\'os \'ut 15-17, 1121 Budapest, Hungary
\and
Max Planck Institute for Astronomy, K\"onigstuhl 17, 69117 Heidelberg, Germany
\and
ELTE E\"otv\"os Lor\'and University, Institute of Physics, P\'azm\'any P\'eter s\'et\'any 1/A, 1117 Budapest, Hungary
\and
Institute of Astronomy and Astrophysics, Academia Sinica, 11F of Astronomy-Mathematics Building, AS/NTU No.1, Sec. 4, Roosevelt Rd, Taipei 10617, Taiwan, ROC
}

\abstract
{}
{Luminosity bursts in young FU Orionis-type stars warm up the surrounding disks of gas and dust, thus inflicting changes on their morphological and chemical composition. In this work, we aim at studying the effects that such bursts may have on the spatial distribution of dust grain sizes and the corresponding spectral index in protoplanetary disks. }
{We use the numerical hydrodynamics code FEOSAD, which simulates the co-evolution of gas, dust, and volatiles in a protoplanetary disk, taking dust growth and back reaction on gas into account. The dependence of the maximum dust size on the water ice mantles is explicitly considered. The burst is initialized by increasing the luminosity of the central star to $100-300~L_\odot$ for a time period of 100~yr. }
{The water snowline shifts during the burst to a larger distance, resulting in the drop of the maximum dust size interior to the snowline position because of more efficient fragmentation of bare grains. After the burst,  the  water  snowline  shifts quickly back  to  its  preburst location followed by renewed dust growth. 
The timescale of dust regrowth after the burst depends on the radial distance so that the dust grains at smaller distances reach the preburst values faster than the dust grains at larger distances. As a result, a broad peak in the radial distribution of the spectral index in the millimeter dust emission develops at $\approx 10$~au, which shifts further out as the disk evolves and dust grains regrow to preburst values at progressively larger distances. This feature is most pronounced in evolved axisymmetric disks rather than in young gravitationally unstable counterparts, { although young disks may still be good candidates if gravitational instability is suppressed. We confirmed our earlier conclusion that spiral arms do not act as strong dust accumulators because of the Stokes number dropping below 0.01 within the arms, but this trend may change in low-turbulence disks.}}
{We argue that, depending on the burst strength and disk conditions, a broad peak in the radial distribution of the spectral index can last for up to several thousand years after the burst has ended and can be used to infer past bursts in otherwise quiescent protostars. The detection of a similar peak in the disk around V883~Ori, an FU Orionis-type star with an unknown eruption date, suggests that such features may be common in the post-outburst objects. }

\authorrunning{Vorobyov et al.}
\titlerunning{Dust in disks of eruptive stars}

\maketitle

\section{Introduction}
Young protostars are known to undergo short episodes of brightening during which their luminosity may increase by several orders of magnitude. These events are called FU Orionis-type eruptions after the first documented prototype and are likely caused by an episodic increase in the mass accretion rate from the circumstellar disk onto the protostar \citep[see][for reviews]{Audard2014,Connelley2018}. In recent years it has become evident that the burst phenomenon may be more widespread than it was originally thought, as the luminosity bursts have also been detected and theoretically confirmed in young massive stars \citep{Caratti2017,Hunter2017,2017MeyerVorobyov,Liu2018}. 

The bursts significantly influence the dynamical and chemical evolution of the circumstellar environment, regardless of the particular mechanism that triggers the burst. The radiative output of the central source heats up the circumstellar disk and envelope, evaporating volatile species and triggering chemical reactions, which otherwise may be dormant \citep{Lee2007,Visser2015,Taquet2016,Rab2017,Molyarova2018,Wiebe2019}. The shift in the water snowline during the burst toward a larger distance than the usual several au makes it possible to test the pile-up and composition of dust grains \citep{Banzatti2015,Schoonenberg2017}. Preferential rather than homogeneous recondensation of water ice on dust nuclei after the burst may promote the formation of gas giants via the streaming instability \citep{Hubbard2017}.

The bursts can also have notable effects on the disk dynamical evolution. An increased luminosity during a century-long burst causes the disk to warm and expand, temporarily stabilizing the disk against gravitational instability and washing out a possible preexisting spiral pattern. A disk contraction that follows expansion can cause disk gravitational fragmentation and giant planet formation in systems that otherwise would show no signs of fragmentation \citep{2020Vorobyov}.  
Therefore, young stars with intermittent luminosity output tend to host circumstellar disks that are more  prone to gravitational fragmentation than their constant luminosity counterparts \citep{2017MercerStamatellos}.  

The effects of the bursts are not limited to the disk chemical and dynamical evolution. They can also affect the stellar evolution, causing excursions in the Herzsprung-Russel diagram if some fraction of accretion energy is absorbed by the star \citep{Kunitomo2017,Elbakyan2019,Meyer2019}. Mass accretion histories with episodic accretion bursts can also explain the age spread in the HR diagrams \citep{Baraffe2012, Jensen2018}. 

Although recent surveys indicate that luminosity variability may be widespread among young stars \citep{ContrerasPena2017}, the classic FU Orionis-type events, for which the pre-burst low-luminosity state is known, have been documented only for a dozen objects. For another few dozens of FUor-like objects the burst onset date is not known, but they show the spectral signatures typical for the classic FUors \citep{Audard2014}. Considering the potential significance of the burst phenomenon for the disk and stellar evolution and the short-lasting nature of the bursts (a few tens to a few hundreds of years), the development of techniques that can help to identify the past bursts has become an important task. 

A promising method is to compare the CO snowline position in the circumstellar envelope with the theoretically predicted one for the current stellar luminosity. A mismatch in the actual and theoretically derived CO snowlines may indicate a luminosity burst in the recent past \citep{Lee2007,Jorgensen2013,Vorobyov2013,Rab2017,Hsieh2019}. However, this method may not work for young T~Tauri stars, in which the envelope has dissipated and CO in the disk is predominantly in the gaseous state \citep{Molyarova2021}. Other less abundant and hence more observationally elusive species have to be attempted \citep{Wiebe2019}. Another method is to search for signatures of intermittent accretion in the spatial structure of protostellar jets \citep{Vorobyov-jets2018}, although the knots may be caused by inhomogeneities in the circumstellar environment \citep{Yirak2008}.

In this work, we explore yet another method that can potentially help to identify the past luminosity bursts, which is based on the analysis of the radial distribution of the spectral index in the millimeter waveband. It was noted that V883~Ori, an FUor-like object with an unknown date of outburst, exhibits a local broad peak in the spectral index at $\approx 42$~au, which was interpreted as the shift in the water snowline during the burst \citep{Cieza2016} assuming the water sublimation temperature of $\approx 105$~K. We show that this feature may last for several thousand years in the disks of eruptive stars, thus signaling the past burst that has long ended. The paper is organized as follows. In Sect.~\ref{model:descript} we provide a brief overview of the numerical model used to simulate the dust evolution in disks of eruptive stars. Sect.~\ref{results} is dedicated to the analysis of our main results. Parameter space study and model caveats are considered in Sects.~\ref{parstudy} and \ref{caveats}. The main conclusions are provided in Sect.~\ref{conclude}.

%Cieza ... Banzatti .. Schnoonenberg ...

%Chemical tracers can be used to constrain the burst mechanisms
%https://ui.adsabs.harvard.edu/abs/2020ApJ...904...78S/abstract

%The magnitudes and durations of luminosity bursts may vary in wide 
\section{Brief description of the numerical model}
\label{model:descript}
Numerical simulations were carried out using the Formation and Evolution of Stars and Disks (FEOSAD) code, which solves the equations of hydrodynamics for the gas and dust disk subsystems in the thin-disk limit.  We refer the reader to  \citet{Molyarova2021} and references therein (in particular, \citet{2018VorobyovAkimkin}) for a complete description. Here we only briefly review the main conceptual parts of the model.

The FEOSAD code computes the co-evolution of gas, dust, and volatiles in a protoplanetary disk on the Eulerian polar grid. The simulations start from a rotating pseudo-disk configuration, which represents a flattened prestellar core that contracts and spins up under its own gravity. The circumstellar disk forms when the first contracting layers of gas hit the centrifugal barrier at the inner computational boundary, which is set equal to 2~au in this work. The disk grows through mass infall from the remaining prestellar core. The material that passes through the inner computational boundary form the central star minus 10\% that are supposed to be ejected with the jets. {This is a rather uncertain number; numerical and theoretical studies provide values ranging from 1\% to 50\% \citep{Wardle1993,Shu1994,Machida2012}, but we have chosen 10\% to be consistent with our previous studies of the accretion burst phenomenon.} Disk self-gravity that takes all disk constituents (gas, dust, volatiles) into account is calculated using the solution of the Poisson integral \citep{BinneyTremaine1987}. The dust-to-gas friction including the back reaction of dust on gas is
calculated using the semi-analytic integration scheme described in \citet{2018Stoyanovskaya}. This scheme was shown to perform well on the Sod shock tube and dusty wave problems. 

\subsection{ Gas component}
The numerical model considers the following main physical processes: viscous torques and heating via the $\alpha_{\rm visc}$-parameterization, disk radiative heating by the central star and circumstellar environment, and disk cooling via dust thermal radiation. The resulting hydrodynamic equations of mass, momentum, and internal energy for the gas component read
\begin{equation}
\label{cont}
\frac{{\partial \Sigma_{\rm g} }}{{\partial t}}   + \nabla_p  \cdot 
\left( \Sigma_{\rm g} \bl{v}_p \right) =0,  
\end{equation}

\begin{eqnarray}
\label{mom}
\frac{\partial}{\partial t} \left( \Sigma_{\rm g} \bl{v}_p \right) +  [\nabla \cdot \left( \Sigma_{\rm
g} \bl{v}_p \otimes \bl{v}_p \right)]_p & =&   - \nabla_p {\cal P}  + \Sigma_{\rm g} \, \bl{g}_p + \nonumber
\\ 
+ (\nabla \cdot \mathbf{\Pi})_p  - \Sigma_{\rm d,gr} \bl{f}_p,
\end{eqnarray}
\begin{equation}
\frac{\partial e}{\partial t} +\nabla_p \cdot \left( e \bl{v}_p \right) = -{\cal P} 
(\nabla_p \cdot \bl{v}_{p}) -\Lambda +\Gamma + 
\left(\nabla \bl{v}\right)_{pp^\prime}:\Pi_{pp^\prime}, 
\label{energ}
\end{equation}
where the subscripts $p$ and $p^\prime$ denote the planar components $(r,\phi)$ in polar coordinates, $\Sigma_{\rm g}$ is the gas mass surface density,  $e$ is the internal energy per surface area, ${\cal P}$ is the vertically integrated gas pressure calculated via the ideal  equation of state as ${\cal P}=(\gamma-1) e$ with $\gamma=7/5$, $f_p$ is the friction force between gas and dust (per dust particle), $\Sigma_{\rm d,gr}$ is the surface density of grown dust, $\bl{v}_{p}=v_r \hat{\bl r}+ v_\phi \hat{\bl \phi}$  is the gas velocity in the disk plane, $\nabla_p=\hat{\bl r} \partial / \partial r + \hat{\bl \phi} r^{-1} \partial / \partial \phi$ is the gradient along the planar coordinates of the disk, and $\bl{g}_{p}=g_r \hat{\bl r} +g_\phi \hat{\bl \phi}$ is the gravitational acceleration in the disk plane that includes the gravity of the central protostar when formed. Turbulent viscosity is taken into account via the viscous stress tensor  $\mathbf{\Pi}$ and the magnitude of kinematic viscosity $\nu=\alpha_{\rm visc} c_{\rm s} H_{\rm g}$ is parameterized using the $\alpha$-prescription of \citet{1973ShakuraSunyaev}, where $c_{\rm s}$ is the sound speed  and $H_{\rm g}$ is the vertical scale height of the gas disk calculated using an assumption of local hydrostatic equilibrium. The expressions for radiative cooling $\Lambda$ and irradiation heating $\Gamma$ (the latter includes the stellar and background irradiation) can be found in \citet{2018VorobyovAkimkin}. The temperature of background radiation is set equal to 15~K.

\subsection{Dust and volatiles}
The dust disk consists of two components: small sub-micron dust with surface density $\Sigma_{\rm d,sm}$ {and sizes in the $5.0\times 10^{-3}-1.0$~$\mu$m range, and grown dust with surface density $\Sigma_{\rm d,gr}$ and sizes from 1.0~$\mu$m and up to a maximum value of $a_{\rm max}$, which depends on the local conditions in the disk.
The boundary between the small and grown dust populations at $a_\ast=1$~$\mu$m is chosen so as to guarantee that small dust is perfectly coupled to gas, while grown dust can dynamically decouple from gas (see Appendix~\ref{coupling} for details). }
The dynamics of small and grown dust grains is described by the following continuity and momentum equations ({ see Appendix~\ref{fluid} for justification of the fluid approach for describing dust dynamics)}
\begin{equation}
\label{contDsmall}
\frac{{\partial \Sigma_{\rm d,sm} }}{{\partial t}}  + \nabla_p  \cdot 
\left( \Sigma_{\rm d,sm} \bl{v}_{ p} \right) = - S(a_{\rm max}),  
\end{equation}
\begin{equation}
\label{contDlarge}
\frac{{\partial \Sigma_{\rm d,gr} }}{{\partial t}}  + \nabla_p  \cdot 
\left( \Sigma_{\rm d,gr} \bl{u}_{ p} \right) = S(a_{\rm max}),  
\end{equation}
\begin{eqnarray}
\label{momDlarge}
\frac{\partial}{\partial t} \left( \Sigma_{\rm d,gr} \bl{u}_{ p} \right) +  \left[\nabla \cdot \left( \Sigma_{\rm
d,gr} \bl{u}_{ p} \otimes \bl{u}_{p} \right)\right]_{ p}  &=&   \Sigma_{\rm d,gr} \, \bl{g}_{p} + \nonumber \\
 + \Sigma_{\rm d,gr} \bl{f}_{ p} + S(a_{\rm max}) \bl{v}_{p},
\end{eqnarray}
where $\bl{u}_{ p}$ are the planar components ($r,\phi$) of the grown dust velocity and  $S(a_{\rm max})$ is the rate of small-to-grown dust conversion per disk surface area (in g~s$^{-1}$~cm$^{-2}$). { The term $\mathbf{f}_p$ is the drag force per unit mass between dust and gas, which is defined as
\begin{equation}
    \mathbf{f}_p = \frac{\mathbf{v}_p - \mathbf{u}_p}{t_{\rm stop}},
    \label{eq:fric}
\end{equation}
where $t_{\rm stop}$ is the stopping time for the Epstein regime expressed as
\begin{equation}
\label{tstop}
    t_{\rm stop} = \frac{a_{\rm max}\, \rho_{\rm s}} {\rho_{\rm g}c_{\rm s}},
\end{equation}
where $\rho_{\rm g}=\Sigma_{\rm g}/(\sqrt{2\pi} H_{\rm g})$ is the volume density of gas and $\rho_{\rm s}$ = 3.0 g cm$^{-3}$ is the material density of dust grains. We checked and confirmed that the Epstein drag regime is not violated in our models, with the minimum Knudsen numbers being in the 300--1700 range depending on the model. 
}

We note that initially the prestellar core has only small dust with the dust-to-gas mass ratio of 0.01, but later part of it is converted to grown dust (see below). Each dust component has the size distribution $N(a)$ described by a simple power-law function { $dN(a)/da = C a^{-p}$ with a fixed exponent $p=3.5$ and a normalisation constant $C$, which can be determined by the mass of dust in each population}. For small dust, the minimum size is $a_{\rm min}=5\times 10^{-7}$\,cm and the maximum size is $a_*=10^{-4}$\,cm. For grown dust, $a_*$ is the minimum size and $a_{\rm max}$ is the dynamic maximum size, which is calculated from the following equation
\begin{equation}
{\partial a_{\rm max} \over \partial t} + ({\bl u}_{\rm p} \cdot \nabla_p ) a_{\rm max} = \cal{D},
\label{dustA}
\end{equation}
where  $\bl{u}_p$ describes the planar components of the grown dust velocity (found from solving the momentum equations for grown dust, see \citet{Molyarova2021}) and  $\nabla_p=\hat{\bl r} \partial / \partial r + \hat{\bl
\phi} r^{-1} \partial / \partial \phi $ is the gradient along the planar coordinates of the disk.  The growth rate $\cal{D}$ accounts for the change in $a_{\rm max}$ due to dust coagulation  and the second term on the left-hand side account for the change of $a_{\rm max}$ due to dust flow through the cell (the left-hand side is the full derivative  of $a_{\rm max}$ over time).  We write the source term $\cal{D}$ as
\begin{equation}
\cal{D}=\frac{\rho_{\rm d} {\it v}_{\rm rel}}{\rho_{\rm s}},
\label{GrowthRateD}
\end{equation}
where  $\rho_{\rm d}$ is the total (small plus grown) dust volume density, $v_{\rm rel}$ is the dust-to-dust collision  velocity, and $\rho_{\rm s}=3.0$~g~cm$^{-3}$ is the material density of dust grains.
{ The total dust volume density is found as 
\begin{equation}
    \rho_{\rm d} = {\Sigma_{\rm d,sm} H_{\rm d} + \Sigma_{\rm d,gr} H_{\rm g} \over \sqrt{2 \pi} H_{\rm d} H_{\rm g} },
    \label{rho:dust}
\end{equation}
where we assumed that the vertical scale height of small dust is equal to that of gas, but grown dust can settle toward the midplane, having defined its scale height $H_{\rm d}$ as a function of the Stokes number $\mathrm{St}$ and $\alpha_{\rm visc}$-parameter \citep{2004Kornet}. The dimensionless Stokes number $\mathrm{St}$ is defined as the product of the Keplerian angular velocity and the stopping time. } The adopted approach is similar to  the monodisperse model of \citet{Stepinski1997} and is described in more detail in \citet{2018VorobyovAkimkin}.
The maximum value of $a_{\rm max}$ is limited  by the fragmentation barrier \citep{Birnstiel2016} defined as
\begin{equation}
 a_{\rm frag}=\frac{2\Sigma_{\rm g}v_{\rm frag}^2}{3\pi\rho_{\rm s}\alpha c_{\rm s}^2},
 \label{afrag}
\end{equation}
where  $v_{\rm frag}$ is the fragmentation velocity of dust grains and $c_{\rm s}$ is the speed of sound. Whenever $a_{\rm max}$ exceeds $a_{\rm frag}$, the growth rate $\cal{D}$ is set to zero. 

The value of fragmentation velocity depends on the ice coating of dust grains and is defined as
\begin{equation}
v_{\rm frag}=
\begin{cases}
15\text{\,m\,s$^{-1}$,} & \text{ if } \Sigma_{\rm ice}/\Sigma_{\rm d,gr} > K , \\
1.5\text{\,m\,s$^{-1}$,} & \text{ if } \Sigma_{\rm ice}/\Sigma_{\rm d,gr} \leq K ,
\end{cases}
\label{eq:vfrag}
\end{equation}
where $\Sigma_{\rm d,gr}$ and $\Sigma_{\rm ice}$ are the surface density of grown dust nuclei (excluding possible ice coating) and the surface density of ice mantles on grown dust grains, respectively. The latter is composed of up to four considered volatile species (see below). The factor $K$ is calculated as in \citet{Molyarova2021}
\begin{equation}
K = \frac{\Sigma_{\rm ice}^{\rm min}}{\Sigma_{\rm d,gr}} ={3 \rho_{\rm ice} \over \rho_{\rm s}} \frac{3 \times 10^{-8} \rm cm}{\sqrt{a_*a_{\rm max}}} ,
\label{eq:threshold}
\end{equation}
where $\Sigma_{\rm ice}^{\rm min}$ is the minimum surface density of ices on grown dust, which is sufficient to cover their surfaces with one monolayer. The material density of ices $\rho_{\rm ice}$ is set equal to 1.0~g~cm$^{-3}$. The values of $v_{\rm frag}=1.5$~m~s$^{-1}$ and $v_{\rm frag}=15$~m~s$^{-1}$ for bare and icy dust grains are chosen based on experimental results \citep{2009ApJ...702.1490W,2015ApJ...798...34G}. 

\begin{table}
\center
\caption{\label{table:2}Frequently used notations}
\begin{tabular}{| c | p{3.5cm} | p{2.7cm}|}
\hline 
quantity  & description   &  reference    \tabularnewline
\hline 
\hline 
%$\cal{P}$ & vertically integrated pressure & ${\cal{P}} = (\gamma - 1) e$ \tabularnewline
%\hline 
%$\gamma$ & adiabatic index &  $\gamma = 7/5$ \tabularnewline
%\hline
$\nu$ & kinematic viscosity &  $\nu = \alpha_\mathrm{visc} c_\mathrm{s} H_\mathrm{g}$ \tabularnewline
\hline 
$H_\mathrm{g}$, $H_\mathrm{d}$ & vertical scale heights of gas and dust &   \tabularnewline
\hline 
%$c_\mathrm{s}$  &    sound speed   &       \tabularnewline
%\hline 
$t_\mathrm{stop}$        &  stopping time         &        see eq.~\eqref{tstop} \tabularnewline
\hline 
$\rho_\mathrm{s}$ & material density of dust &  $\rho_\mathrm{s}$ = 3 g cm$^{-3}$ \tabularnewline
\hline 
$\mathrm{St}$ & Stokes number & $\mathrm{St} = t_\mathrm{stop} \Omega $ \tabularnewline 
\hline 
$a_\mathrm{min}$    & minimal size of small dust & $a_{\rm min} = 5 \cdot 10^{-7}$ cm       \tabularnewline
\hline 
$a_{*}$ & small and grown dust borderline & $a_{*} = 10^{-4}$ cm \tabularnewline
\hline 
$a_\mathrm{max}$ & maximal size of grown dust & see eq. \eqref{dustA} \tabularnewline
\hline 
$\cal{D}$  & collisional growth rate& see eq. \eqref{GrowthRateD}       \tabularnewline
\hline 
$a_\mathrm{frag}$ & fragmentational barrier & see eq. \eqref{afrag}       \tabularnewline
\hline 
$v_\mathrm{frag}$ & fragmentation velocity & see eq.~(\ref{eq:vfrag}) \tabularnewline 
\hline 
$p$ & slope of the grain size distribution & $p = 3.5$ \tabularnewline
\hline 
$\alpha_\mathrm{visc}$ & viscosity parameter & $\alpha_\mathrm{visc} = 10^{-2}$ and eq.~\eqref{alphaV} \tabularnewline
\hline 
$\zeta_\mathrm{d2g}$ & dust-to-gas mass ratio & $\zeta_\mathrm{d2g} = {\Sigma_\mathrm{d,tot}} / {\Sigma_\mathrm{g}}$ \tabularnewline
\hline 
$\kappa_{\nu}$ & opacity coefficient & cm$^2$ g$^{-1}$   \tabularnewline
\hline 
$\alpha \left( \lambda_{1}, \lambda_{2} \right)$ & spectral index & see eq.~\eqref{alpha_spectral} \tabularnewline
%4 &  300 & 0.01 &  390 & 0.5:5 \tabularnewline
\hline 
\end{tabular}
\end{table}

\begin{figure*}
\begin{centering}
\includegraphics[width=0.66\columnwidth]{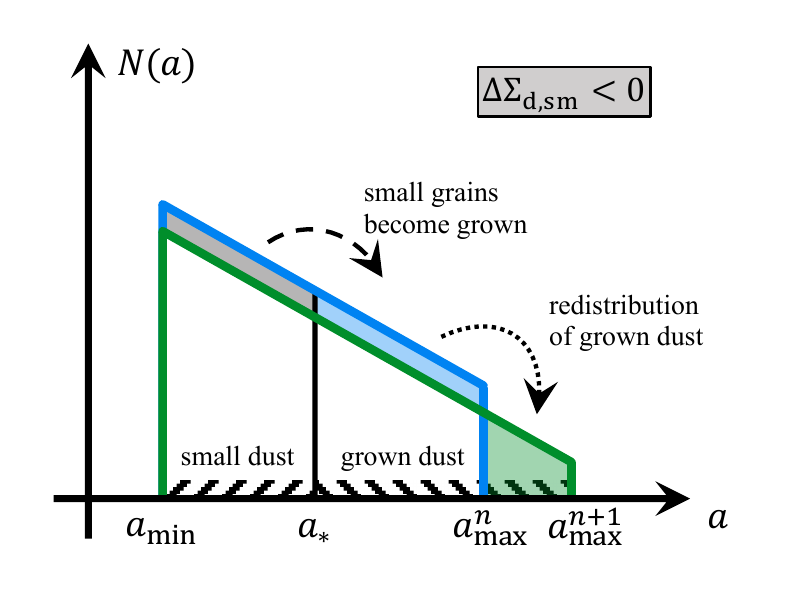}
\includegraphics[width=0.66\columnwidth]{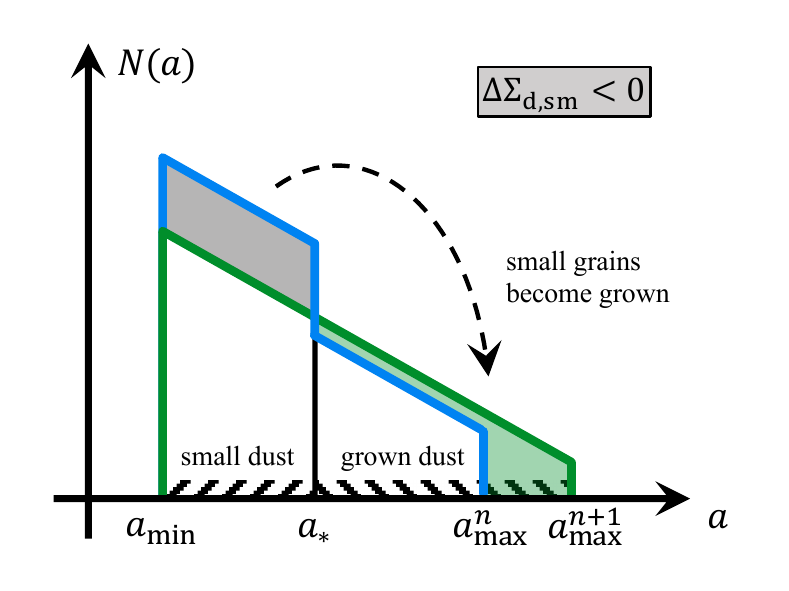}
\includegraphics[width=0.66\columnwidth]{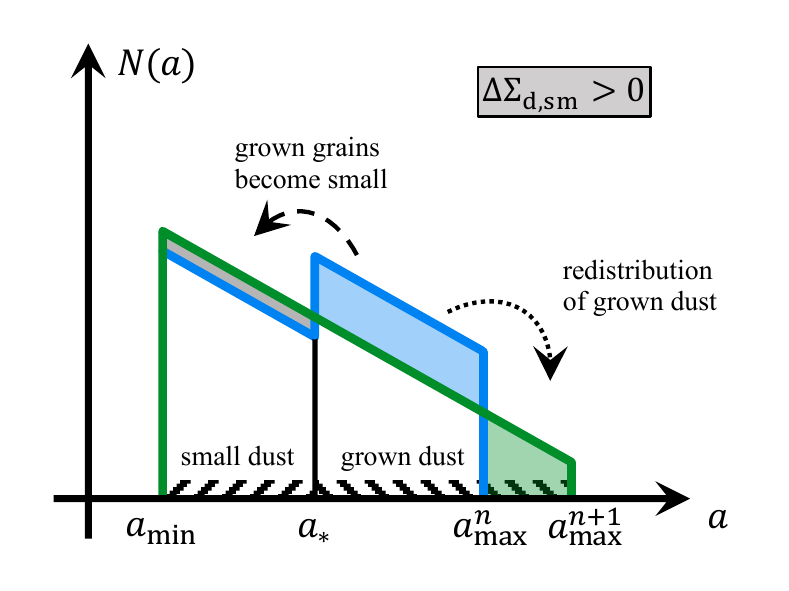}
\par\end{centering}
\caption{\label{fig:0}  Illustration of dust evolution in the two-sized dust population scheme. Three possible examples of dust size distribution at the current time step (blue lines, $a_{\rm max}^n$) and the next time step (green lines, $a_{\rm max}^{n+1}$ ) are illustrated. {\bf Left:} continuous distribution remain continuous during dust growth;  {\bf center:} discontinuous distribution with an excess of small dust turns into a continuous one; {\bf right:} discontinuous distribution with an excess of grown dust turns into a continuous one. The change in small dust surface density $\Delta \Sigma_{\rm d,sm}$ is shaded with grey, while added and removed grown dust is shaded by green and blue colors, respectively.}
\end{figure*}

The dynamics and phase transitions of volatile species are calculated
using the following equations
\begin{eqnarray}
\frac{\partial \Sigma_{s}^{\rm gas}}{\partial t} + ({\bl v}_{\rm p} \cdot \nabla_p ) \Sigma_{s}^{\rm gas} &=&-\lambda_s\Sigma_{s}^{\rm gas}+\eta_s^{\rm sm}+\eta_s^{\rm gr},\label{eq:sig1}\\
\frac{\partial \Sigma_{s}^{\rm sm}}{\partial t} + ({\bl v}_{\rm p} \cdot \nabla_p ) \Sigma_{s}^{\rm sm} &=&\lambda_s^{\rm sm}\Sigma_{s}^{\rm gas}-\eta^{sm}_{\rm s},\label{eq:sig2}\\
\frac{\partial\Sigma_{s}^{\rm gr}}{\partial t} + ({\bl u}_{\rm p} \cdot \nabla_p ) \Sigma_{s}^{\rm gr}  &=&\lambda_s^{\rm gr}\Sigma_{s}^{\rm gas}-\eta_s^{\rm gr},\label{eq:sig3}
\end{eqnarray}
where $\Sigma_{s}^{\rm gas}$, $\Sigma_{s}^{\rm sm}$, and $\Sigma_s^{\rm g}$ are the surface densities of four volatile species (CO, CO$_2$, H$_2$O, and CH$_4$) in the gas phase and on the surfaces of small and grown dust, respectively.
The  adsorption rates onto small and grown dust grains are
$\lambda_s^{\rm sm}$ and $\lambda_s^{\rm gr}$ (s$^{-1}$), and  $\lambda_s=\lambda_s^{\rm sm}+\lambda_s^{\rm gr}$. We also define $\eta_s=\eta_s^{\rm sm}+\eta_s^{\rm gr}$ (g\,cm$^{-2}$\,s$^{-1}$), where $\eta_s^{\rm sm}$ and $\eta_s^{\rm gr}$ are the desorption rates from small and grown dust populations, respectively.
The model includes both thermal desorption and photo-desorption, although in the context of this study thermal desorption due to heating by the outburst is dominant.
For the definition of the adsorption and desorption rates we refer the reader to \citet{Molyarova2021}, noting here that we use the zero-order desorption model. For the initial composition of volatiles we use the ice abundances measured with \textit{Spitzer} in protostellar cores \citep{Oberg2011}. The binding energies of the considered volatile species were taken from \citet{Aikawa1996,Bisschop2006,Brown2007,Noble2012}. The possible chemical transformations of the considered species are not taken into account as this requires a more complex chemodynamical treatment, which is beyond the scope of this study. 

\subsection{Small to grown dust conversion}

In the FEOSAD code, small dust can be converted to grown dust as the disk evolves and $a_{\rm max}$ increases. The conversion scheme assumes that the total (small plus grown) dust mass in each numerical cell is conserved when dust is converted from the small to grown component. 
{ Furthermore, if the two-sized dust distribution becomes discontinuous at the boundary between the two populations ($a_\ast$), which may occur owing to differential dust drift of small and grown dust, then we assume that dust growth smooths out this discontinuity (see Fig.~\ref{fig:0}).  Physically this assumption corresponds to the dominant role of dust collisional evolution over dust flow in setting the fixed and continuous slope of the dust size distribution across two dust populations. }
%In addition, we assume that the slope of the dust size distribution $p=3.5$ does not change due to dust growth. 
These stipulations allowed us to derive a simple dust growth method  with the rate of small-to-grown dust conversion $\Delta \Sigma_{\rm d,sm}=-\Delta \Sigma_{\rm d,gr}=\Sigma_{\rm d,sm}^{n+1}-\Sigma_{\rm d,sm}^{n}$ expressed as \citep[see][for details]{Molyarova2021}
\begin{equation}
\label{final}
    \Delta\Sigma_{\mathrm{d,sm}} = 
    \frac
    {
    \Sigma_{\rm d,gr}^n I_1  - 
    \Sigma_{\rm d,sm}^n I_2
    }
    {
    I_3
    },
\end{equation}
where $\Sigma_{\rm g,sm}$ is the surface density of small dust and the integrals $I_1$, $I_2$, and $I_3$ are defined as
\begin{eqnarray}
   I_1= \int_{a_{\rm min}}^{a_*} a^{3-\mathrm{p}}da,\nonumber
   \\
   I_2=\int_{a_*}^{a_{\mathrm{max}}^{n+1}} a^{3-\mathrm{p}}da.\nonumber \\
    I_3= \int_{a_{\mathrm{min}}}^{a_{\rm max}^{n+1}} a^{3-\mathrm{p}}da.
\end{eqnarray}
Here, the superscripts $n$ and $n+1$ correspond to the current and next time step. The rate of small-to-grown dust conversion in Equations~(\ref{contDsmall})--(\ref{momDlarge}) during one time step $\Delta t$ is then written as
\begin{equation}
S(a_{\rm max})=-\Delta \Sigma_{\rm d,sm}/\Delta t.
\end{equation}

We note that formally our scheme also allows the back conversion of grown to small dust when $a^{\rm n+1}_{\rm max}<a^{\rm n}_{\rm max}$, namely, when the maximum dust size decreases with time. This is feasible if the disk conditions evolve in such a manner that the fragmentation barrier decreases in some disk locations, for instance, because of a sudden heating event by a luminosity burst. In this case, the maximum size of dust grains is reset to $a_{\rm frag}$, meaning that part of the grown dust reservoir is converted back to small dust. In the present study, this process is assumed to occur on timescales that are much faster than dust growth due to mutual dust-to-dust collisions. This is feasible if dust aggregates are glued together by icy mantles, as was already suggested from numerical simulations of dust growth with freeze-out and sublimation of volatiles by \citet{Molyarova2021},  and the burst melts these mantles because of the rising temperature.  We plan to explore the limitations of this approach in a follow-up study. 
We do not take the possible dust sublimation at high temperatures into account (because the temperature in our models never exceeds 1500~K in the computational domain $\ge~2$~au) and neglect dust diffusion associated with turbulence on our timescales of interest.

{ We note that the span in the grain sizes of the grown dust component (from $a_\ast$ to $a_{\rm max}$)  may be substantial. However, we calculate the stopping time in Eq.~(\ref{tstop}) using $a_{\rm max}$, and not an effective dust size that is averaged over the entire grain size distribution of the grown dust component. This means that our model tracks the dynamics of the upper end of the dust size distribution where most of the dust mass reservoir is located for a power index of $p=3.5$.}
{ Finally, we emphasize that our disk model is not razor-thin and we account for the disk vertical extent  by computing the vertical scale heights of gas and dust ($H_{\rm g}$ and $H_{\rm d}$) in each computational cell. These quantities are further used in the calculations of the heating function $\Gamma$ (Eq.~\ref{energ}) when accounting for the stellar radiation flux intercepted by the disk surface, in the kinematic viscosity $\nu$, and in the dust volume density (Eq.~\ref{GrowthRateD}) when considering possible vertical settling of grown dust. For the reader's convenience, Table~\ref{table:2} summarized the most often used notations in this work.}

\subsection{ Initializing the burst}
To set up a luminosity burst, we note that the inner 2~au of the disk are not dynamically resolved in our hydrodynamic model and are replaced with the sink cell. We assume that a luminosity burst can be triggered by the development of the magnetorotational instability or thermal instability in these innermost disk regions \citep[e.g.][]{Bae2014,Kadam2020} or even convective instability \citep{Pavlyuchenkov2020,Maksimova2020}. We account for this effect by  artificially increasing the stellar luminosity to 300~$L_\odot$, a value that is within the limits inferred for FUors \citep{Audard2014}. This corresponds to approximately factors of 75 and 140 increase in the preburst luminosity at $t=90$~kyr (model~1) and $t=290$~kyr (model~2), respectively. 
We note that the chosen evolutionary times correspond roughly to the embedded and early T~Tauri phases of disk evolution, for which FUor eruptions have been confirmed through observations and numerical modeling \citep{Quanz2007,Bae2014,VorobyovBasu2015}.
The duration of the burst is set equal to 100 yr for all models, which approaches the longest confirmed burst in FU Orionis and is likely shorter than the burst in V883 Ori \citep{2020Vorobyov}.

\begin{figure*}
\begin{centering}
%\resizebox{\hsize}{!}{\includegraphics{figure3b.jpg}}
\includegraphics[width=\textwidth]{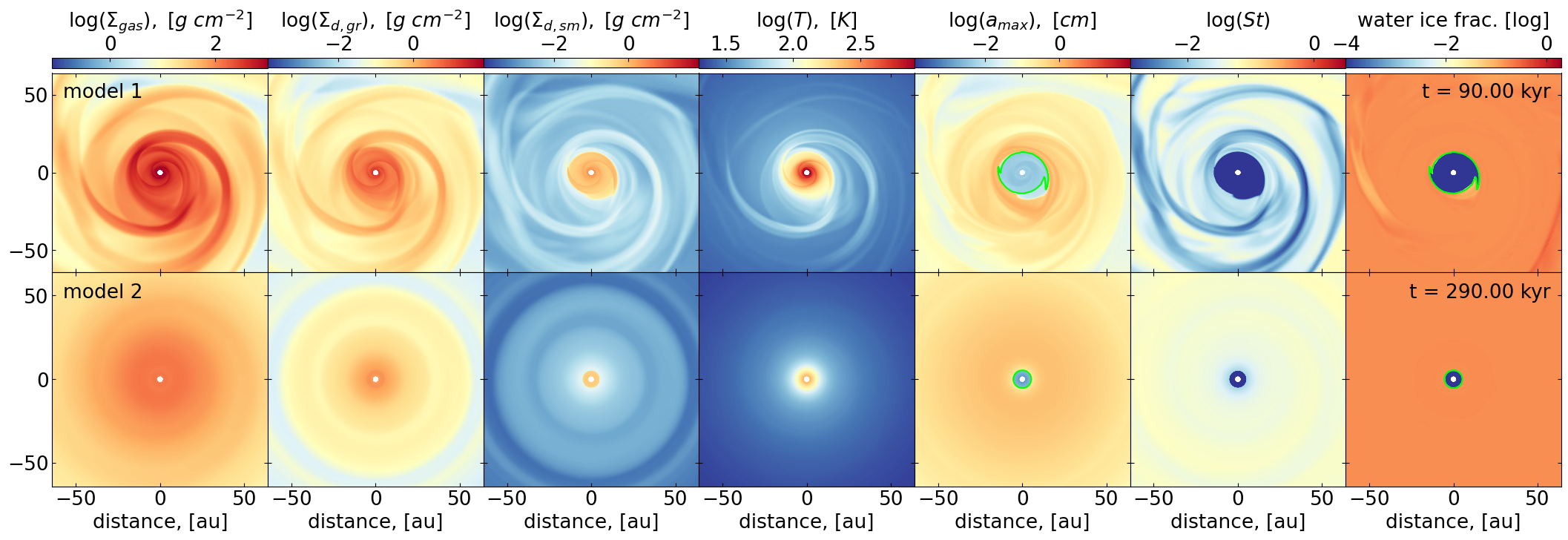} 
\par \end{centering}
\caption{Spatial distributions of various disk characteristics just before the onset of the burst at $t=90$~kyr (model~1, top row) and $t=290$~kyr (model~2, bottom row) after the disk formation instance. Columns from left to right represent the gas surface density, grown dust surface density, small dust surface density, temperature, maximum dust size, Stokes number, and the mass fraction of water ice on dust grains $\xi_{\rm H_2O\#}= (\Sigma_{\rm d,gr}^{\rm H_2O\#} + \Sigma_{\rm d,sm}^{\rm H_2O\#}) / \Sigma_{\rm d,tot}$, respectively. The green lines outline the position of the H$_2$O snowline.}
\label{fig:init}
\end{figure*}

\section{Results}
In this section, we consider the effects of a luminosity burst on various disk characteristics and, in particular, on the spatial distribution of dust grain sizes. A special attention was paid to the time evolution of the spectral index at the millimeter wavelength band after the burst.

\label{results}
\subsection{Preburst disk configuration}
The long-term evolution of gas, dust, and volatiles in a protoplanetary disk  was studied in detail in  \citet{Molyarova2021}. These authors employed the FEOSAD code and followed the disk dynamics starting from its formation till an age of 0.5~Myr. 
Two values of the viscous $\alpha_{\rm visc}$-parameter were considered: $\alpha_{\rm visc}=0.01$ and $\alpha_{\rm visc}=10^{-4}$. The disk simulations of \citet{Molyarova2021} are best suited for studying the effects of the burst because these models take the dynamics and phase transformations of volatile species self consistently into account, including H$_2$O that is of particular interest for us  (see Eqs.~\ref{eq:sig1}-\ref{eq:sig3}). 
We take the preburst disk configurations from \citet{Molyarova2021} and consider the model with the viscous $\alpha_{\rm visc}$-value set equal to a spatially and temporally constant value of 0.01. This implies efficient mass and angular momentum transport throughout the entire disk provided either by the magnetorotational instability. A model with an adaptive $\alpha_{\rm visc}$-value will be considered within the framework of the parameter study in Sect.~\ref{parstudy}. Furthermore, we select two time instances that are meant to represent different disk evolutionary stages: an early gravitationally unstable stage and a more evolved gravitationally stable stage. These two models will be referred to in the following text as model~1 and model~2. Figure~\ref{fig:init} presents the main disk characteristics in the inner $120\times120$~au$^2$ box at t=$90$~kyr (model~1, top row) and $t=290$~kyr (model~2, bottom row) from the disk formation instance (which is 60~kyr after the onset of cloud core collapse). Clearly, the disk at $t=90$~kyr is gravitationally unstable and exhibits a spiral pattern, whereas at $t=290$~kyr the disk is gravitationally stable and, as a consequence, nearly axisymmetric (see Appendix~\ref{Qparam} for details).

Initially the entire dust mass reservoir in the prestellar core is in the form of small sub-micron dust grains. At $t=90$~kyr the surface density of grown dust exceeds that of small dust, meaning that efficient dust growth occurs as early as in the disk formation stage \citep[see also]{Chiang2012,2018VorobyovAkimkin,Galametz2019}, but in some objects perhaps not earlier than the late Class 0 stage \citep{Li2017,Bliu2021}. The spiral pattern is prominent in both gas and dust. We define the water snowline as  a position in the disk where the surface density of water ice on both dust populations is equal to the surface density of water in the gas phase. The position of the water snowline may differ with the distance from midplane and the current definition is more appropriate for the water snowline in the disk midplane. 
Figure~\ref{fig:init} indicates that the water snowline is located at a notably larger distance from the star in a younger disk, namely, at $\approx 13.6$~au in model~1 (after azimuthal averaging)  and $\approx 5.4$~au in model~2. This is caused by a higher disk temperature in the early disk evolution stage owing to stronger viscous heating and more luminous younger star ($L_\ast=4.1~L_\odot$ at $t=90$~kyr vs. $L_\ast=2.1~L_\odot$ at $t=290$~kyr).
The maximum size of dust grains decreases notably inside the water snow line because of the drop in the fragmentation barrier $a_{\rm frag}$ for bare dust particles (see Eq.~\ref{eq:vfrag}). Here and further in the text, we define the snowline position in the disk midplane as the  spatial location where  $\Sigma^{\rm sm}_{\rm s} + \Sigma^{\rm gr}_{\rm s} = \Sigma^{\rm gas}_{\rm s}$. Because water is the main species responsible for the changes in $a_{\rm frag}$ in our model, we leave the other volatile species (CO, CO$_2$, and CH$_4$) from consideration.

\begin{figure}
\begin{centering}
\resizebox{\hsize}{!}{\includegraphics{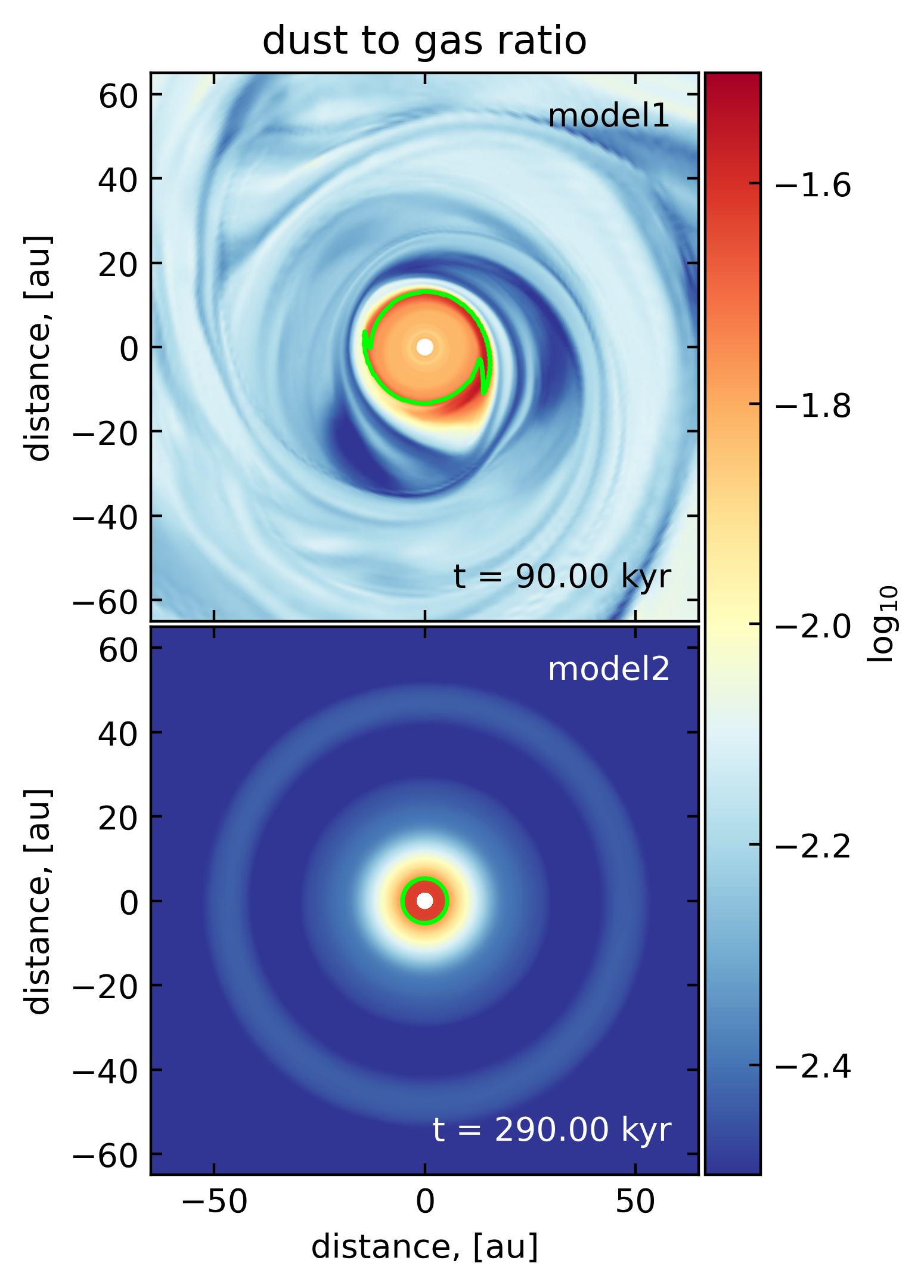}}
\par \end{centering}
\caption{Spatial distribution of the total dust-to-gas mass ratio in the young and evolved disks (top and bottom panels, respectively). The green lines outline the position of the H$_2$O snowline. } 
\label{fig:d2g}
\end{figure}

\begin{figure}
\includegraphics[width=\linewidth]{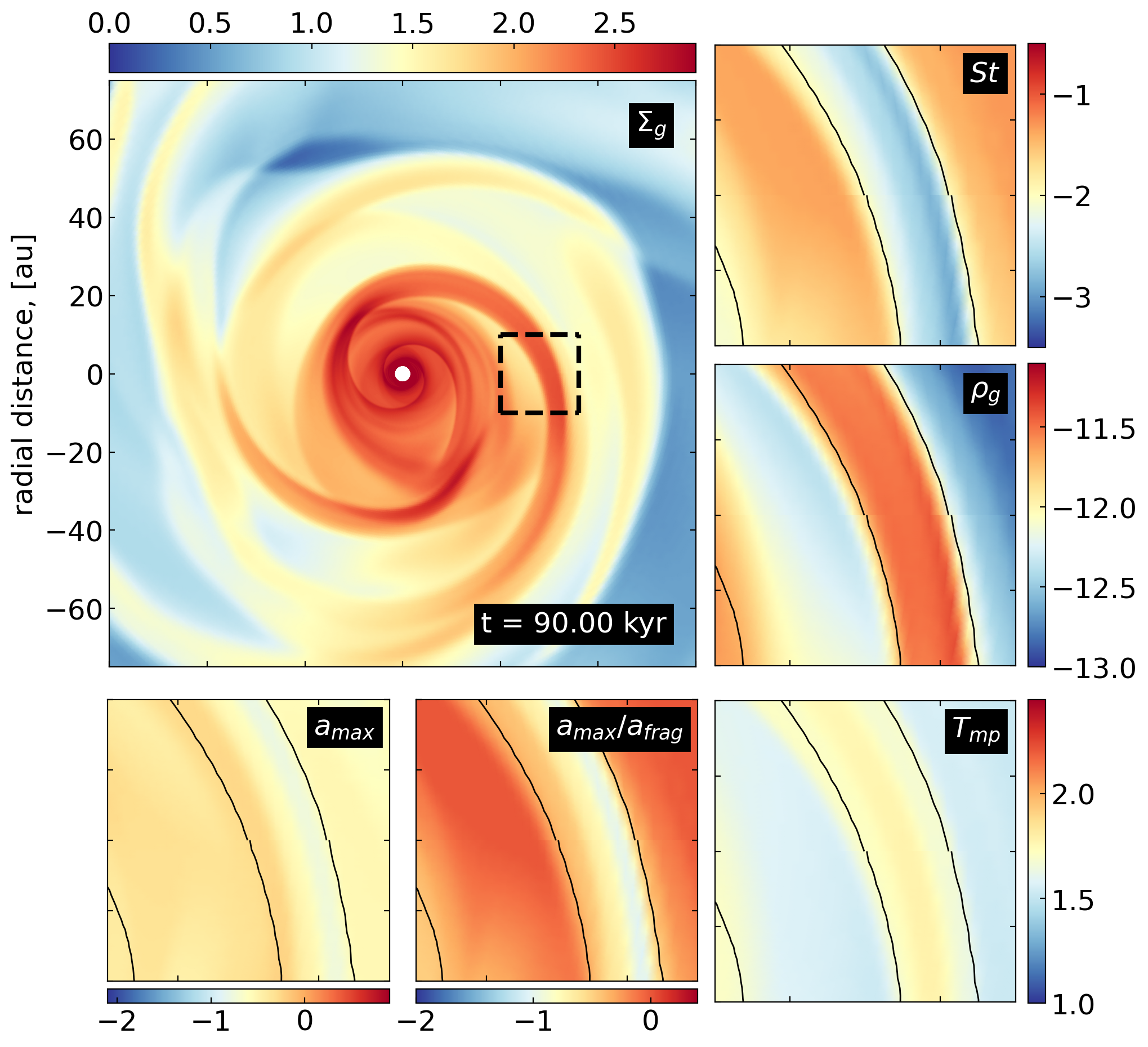}
\caption{Analysis of disk characteristics in the vicinity of the spiral arm in model~1 at t=90~kyr. The analyzed element of the arm is outlined with the black dashed line in the upper-left panel. The other smaller panels show the Stokes number $\mathrm{St}$, volume density of gas $\rho_{\rm g}$, gas temperature in the midplane $T_{\rm mp}$, ratio of the maximum dust size to fragmentation barrier $a_{\rm max}/a_{\rm frag}$, and maximum size of dust grains $a_{\rm max}$. The solid curves outline the disk regions with $\mathrm{St}<0.01$. }
\label{fig:stokes}
\end{figure}

Figure~\ref{fig:d2g} presents the spatial distribution of the total dust-to-gas mass ratio $\zeta_{\rm d2g}=(\Sigma_{\rm d,sm}+\Sigma_{\rm d,gr})/\Sigma_{\rm g}$, at $t=90$~kyr and $t=290$~kyr.
Notable deviations from the initial value of 0.01 are evident in the disk already in the early gravitationally unstable evolution stage \citep[see also][]{VorobyovElbakyan2019,Commercon2020}. In general, the disk regions that are located appreciably outside the water snowline are depleted in dust, while the disk regions inside and in the vicinity of the snowline are enriched in dust. This effect is known as the ``traffic-jam'', { which is a non-uniform dust drift that may occur owing to pressure gradient variations in the disk or spatially different sizes of dust particles and resulting drift velocities} \citep[e.g.,][]{Birnstiel2010,Pinilla2016,Molyarova2021}, and it was suggested to promote planet formation \citep[e.g.][]{Zhang2015}.

Interestingly, there is no significant overabundance of dust in the spiral arms compared to the inter-arm regions. It should be noted that the spiral structure in our models has an agile character, continuously changing its pattern while the arms are forming, winding up, and re-forming again. This acts against accumulation of dust. As was shown by \citet{2018VorobyovAkimkin}, mild dust accumulation is only possible near corotation of the spiral pattern with the disk where the dust drift timescale is shorter than the dynamical timescale of dust flow through the spiral arm. { Another reason for that can be seen in the spatial distribution of the Stokes number shown in Figure~\ref{fig:stokes} -- it falls below 0.01 in the spiral arms, meaning that these structures become inefficient dust accumulators (see justification in Appendix~\ref{coupling}).  As Figure~\ref{fig:stokes} demonstrates, a decrease in the Stokes number is mainly caused by an increase of the gas volume density  by about an order of magnitude compared to the inter-arm regions.  The sound speed and the gas temperature also increase owing to compressional heating in the optically thick spiral arms but by a smaller factor. Interestingly, the fragmentation barrier $a_{\rm frag}$ increases in the arms because an increase in the gas density overweighs that of the gas temperature (see~Eq.~\ref{afrag}). As a result, dust starts growing  and $a_{\rm max}$ increasing in the arms to reach the new upper limit imposed by $a_{\rm frag}$, which also leads to an increase in the Stokes number. However, the growth timescale may be shorter than the time that dust grains spend inside the spiral density waves as they orbit the central star. The effect may become notable only in the vicinity of the corotation radius between the gas spiral pattern and the dust disk \citep{2018VorobyovAkimkin}. All these positive and negative feedback loops deserve a focused study in a follow-up paper with different values of the $\alpha_{\rm visc}$ parameter.}

We note that several previous studies have found the opposite, namely that the spiral arms are efficient dust traps \citep[e.g.,][]{2004RiceLodato,Boss2020}. In particular, \citet{Boss2020} suggested that the inefficient dust accumulation is caused by the two-dimensional nature of disk simulations in the FEOSAD code. We note, however, that these works have adopted dust particles of constant size and/or constant Stokes number. In our model, the maximum dust size and Stokes numbers are spatially and temporally variable and adapt to the local conditions in the spiral arms. These conditions (low Stokes number) are not favorable for dust accumulation, meaning that the effect is due to a more self-consistent nature of our model rather than due to the lower dimesionality of numerical simulations. Nevertheless, it should be confirmed with full three-dimensional models. { We note that decreasing the $\alpha_{\rm visc}$-parameter (recall that our fiducial model has $\alpha_{\rm visc}=10^{-2}$) and thus raising the fragmentation barrier may favour dust growth and may increase the Stokes number. We plan to investigate if spiral arms in low-$\alpha_{\rm visc}$ disks can serve as dust traps in a follow-up study. }

\subsection{Effects of a luminosity burst}

 \begin{figure*}
\begin{centering}
%\resizebox{\hsize}{!}{\includegraphics{figure3b.jpg}}
\includegraphics[width=\textwidth]{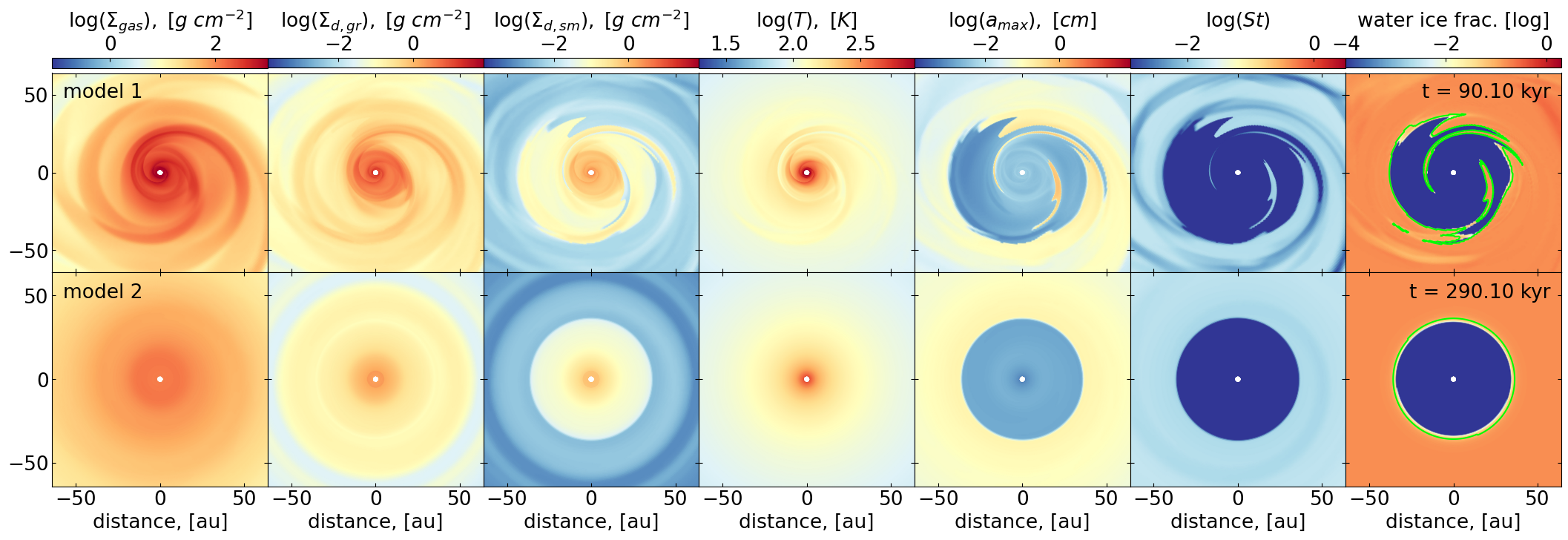} 
\par \end{centering}
\caption{Similar to Fig.~\ref{fig:init} but at the end of the $300~L_\odot$ burst.} 
\label{fig:endburst}
\end{figure*}

Figure~\ref{fig:endburst} displays the disk characteristics in models~1 and 2 at the end of the $300~L_\odot$ luminosity burst at $t=90.1$~kyr and $t=290.1$~kyr, respectively. The adopted duration of the burst is short compared to the dynamical time and cannot affect notably the shape of the spiral pattern \citep{2020Vorobyov}.  Nevertheless, the burst has an appreciable impact on the gas temperature, maximum dust size, and the position of the water snowline, because the timescale for stellar radiative heating is shorter than the disk orbital period beyond 10--20~au (see Appendix~\ref{Append:coolheat}). 
In particular, the burst raises the gas temperature in the bulk of the disk except for the innermost disk regions where the disk heating is mostly provided by turbulent viscosity and PdV work. The water snowline shifts further out to $\approx 37-39$~au in both models, resulting in a decrease of $a_{\rm max}$ and the Stokes number in the disk regions interior to the snowline. Conversely, the surface density of small dust  increases in these regions.
We note that the water snowline deviates notably from a circular shape in the young gravitationally unstable disk (model~1), which is a consequence of the heating by spiral density waves \citep[for details see][]{Molyarova2021}, an effect that can only be explored in multi-dimensional hydrodynamic simulations.

\begin{figure*}
\begin{centering}
%\resizebox{\hsize}{!}{\includegraphics{figure3b.jpg}}
\includegraphics[width=\textwidth]{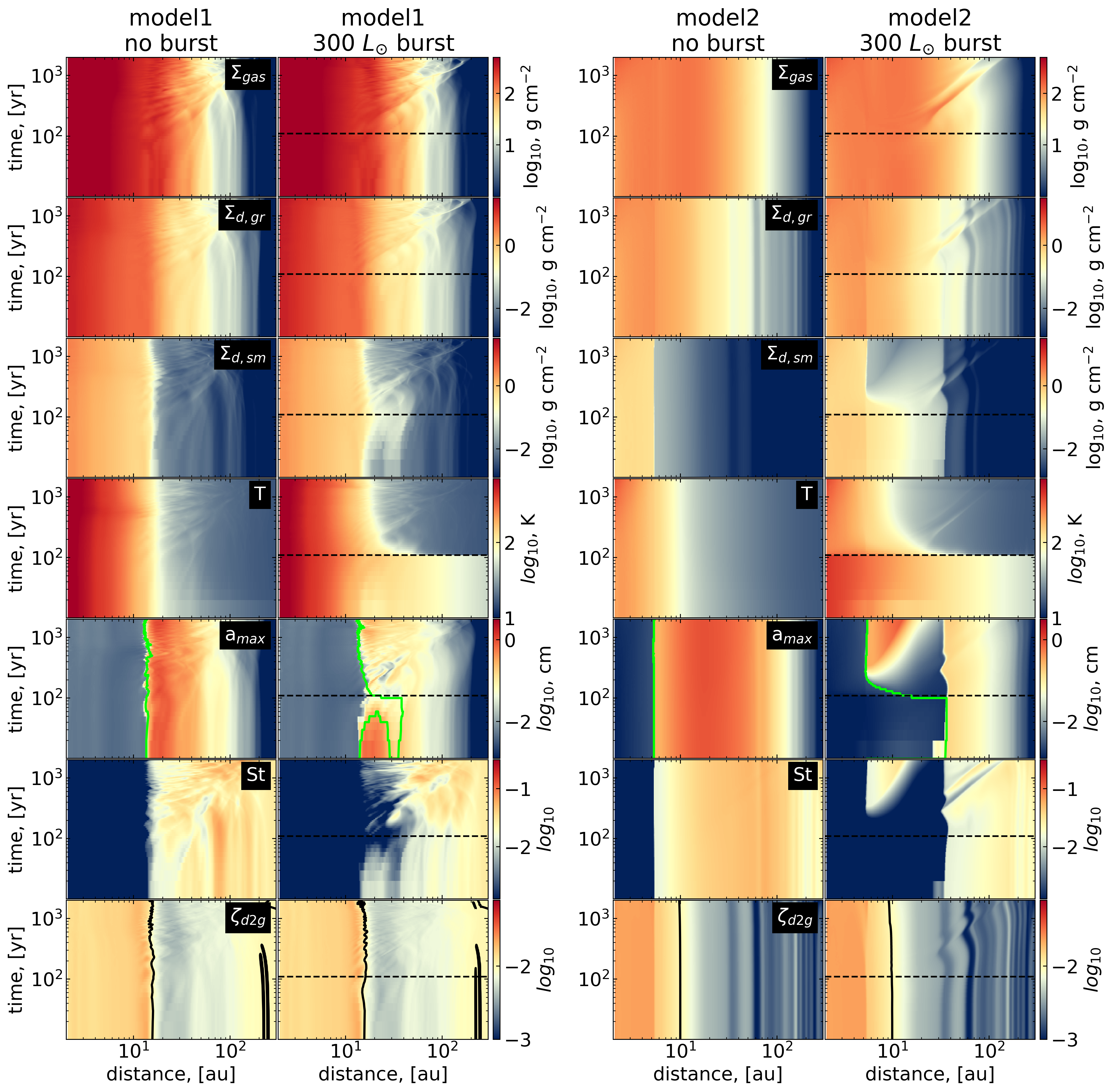} 
\par \end{centering}
\caption{Temporal evolution of the azimuthally-averaged disk characteristics for model~1 with the early ignition of the burst (left-hand side duplet of columns) and model~2 with the late ignition (right-hand side duplet). The time is counted from $t=90$~kyr in model~1 and $t=290$~kyr in model~2. Each column in the duplets corresponds to the case without burst (left column) and with a burst of $L = 300~L_{\odot}$ (right column). The columns from top to bottom show the gas surface density, grown dust surface density, small dust surface density, temperature, maximal size of dust grains, Stokes number, and dust-to-gas mass ratio, respectively. The horizontal black dashed lines indicate the terminal time of the burst. The thick green curves in the fifth row outline the position of the water snow line. The black solid curves in the bottom row identify the disk loci with the total dust to gas ratio of 0.01 for convenience.} 
\label{fig:4}
\end{figure*}

Figure~\ref{fig:4} presents the azimuthally averaged disk characteristics over a time period of 2.0~kyr for a young disk in model~1 (two left-hand side columns) and evolved disk in model~2 (two right-hand side columns).   The end of the burst is marked by the dashed horizontal line. For comparison purposes we also show the disk characteristics over the same time period in the models without the burst. Clearly, the bursts influence notably the disk temperature, surface density of small dust, and dust size distribution. We note that for the adopted dust size slope of $p=3.5$ most of the dust mass is concentrated in the grown dust component. This is why the conversion of small to grown dust component has a greater impact on the surface density of small dust. The position of the water snowline predictably shifts to a larger distance, but its time evolution is different in the young and evolved disks. In the evolved disk the snowline moves quickly to a distance of $\approx 37$~au (which is close to an estimated distance of the water snowline in V883~Ori at 42~au \citet{Cieza2016}), whereas in the young disk a snow ``ring'' develops at $\approx 20$~au, which disappears only closer the end of the burst. This ring is a result of azimuthal averaging of the non-axisymmetric disk in model~1 and it represents in fact the elements of spiral arms that are optically thicker than their immediate surroundings (see the last column in Fig.~\ref{fig:endburst}). As a result, it takes a longer time for the burst to warm up the spirals to the water ice sublimation temperature.  
 
Most interestingly, the effect of the burst is retained in certain dust characteristics long after the burst has ended.  Although the disk temperature returns to the preburst state over just a few years (because of fast cooling timescales), the surface density of small dust and the maximum size of dust grains return to the preburst values during a much longer time period. This is particularly true for the evolved disk in model~2. Let us consider the water snowline, which divides the disk into two regions: the inner part dominated by bare grains with a  maximum size of a few tens of microns and the outer part with dust grains reaching a size of several millimeters. During the burst, the water snowline shifts to a larger distance and the maximum size of dust grains drops following the decrease in $a_{\rm frag}$ for bare dust grains. As the burst ends, the water snowline quickly returns to the preburst position. However, the dust size distribution reacts to the end of the burst much slower and returns to the preburst state on timescales of thousands of years. The same is true for the conversion timescales of small to grown dust after the burst.

\begin{figure}
\includegraphics[width=\linewidth]{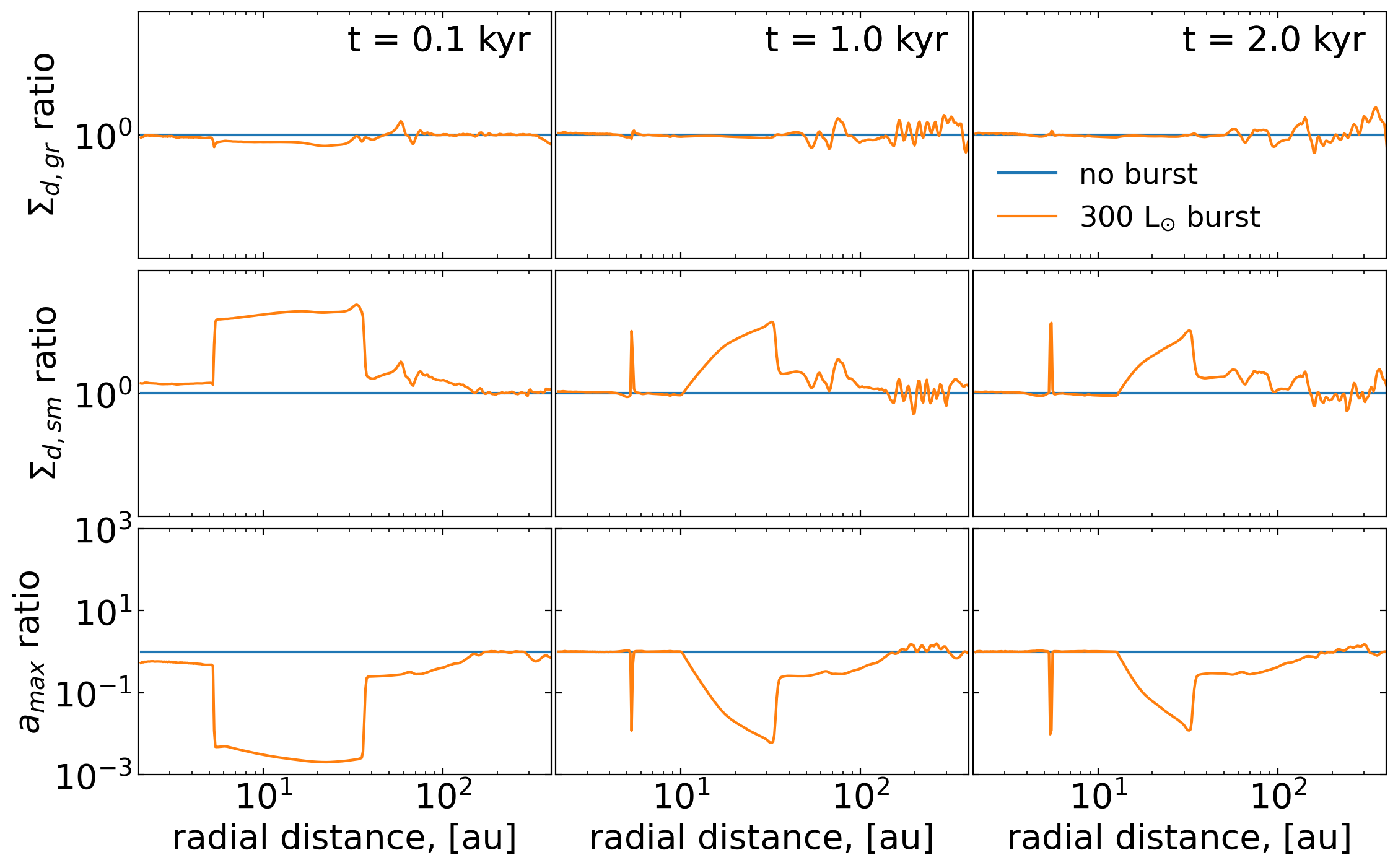}
\caption{Ratios of dust characteristics at several evolution times after the burst onset to those of the model without burst (orange curves). The columns from left to right correspond to the evolution times $t=0.1$~kyr, $t=1$~kyr and $t=2$~kyr after the burst onset in model~2. Rows from top to bottom show the ratios of the azimuthally averaged grown dust surface density, small dust surface density, and maximum grain size. The blue lines show the unity ratios for convenience.}
\label{fig:ratios}
\end{figure}

We illustrate this effect in Figure~\ref{fig:ratios}, which presents the ratios of the azimuthally averaged grown dust surface density, small dust surface density, and maximum dust size in the burst model~2 to their respective values in the nonburst model.  Three distinct evolution times are shown: 100~yr, 1~kyr, and 2~kyr after the end of the burst. Clearly, the burst has a notable impact on $\overline{\Sigma}_{\rm d,sm}$ and $\overline{a}_{\rm max}$, increasing the former and lowering the latter. {Here and further in the text, the overbar sign over variables  means azimuthal averaging at a specific radial distance}. The effects are strongest at the end of the burst but are noticeable even 2.0~kyr after the burst. We note that a sharp spike at about 5~au is caused by the mismatch between the snowline positions in the models without and with the burst. The burst causes the disk to expand \citep{2020Vorobyov} and the burst model does not recover the exact same position of the snowline in the postburst stage because of the changing physical conditions in the disk.

\begin{figure}
\begin{centering}
%\resizebox{\hsize}{!}{\includegraphics{figure3b.jpg}}
\includegraphics[width=\linewidth]{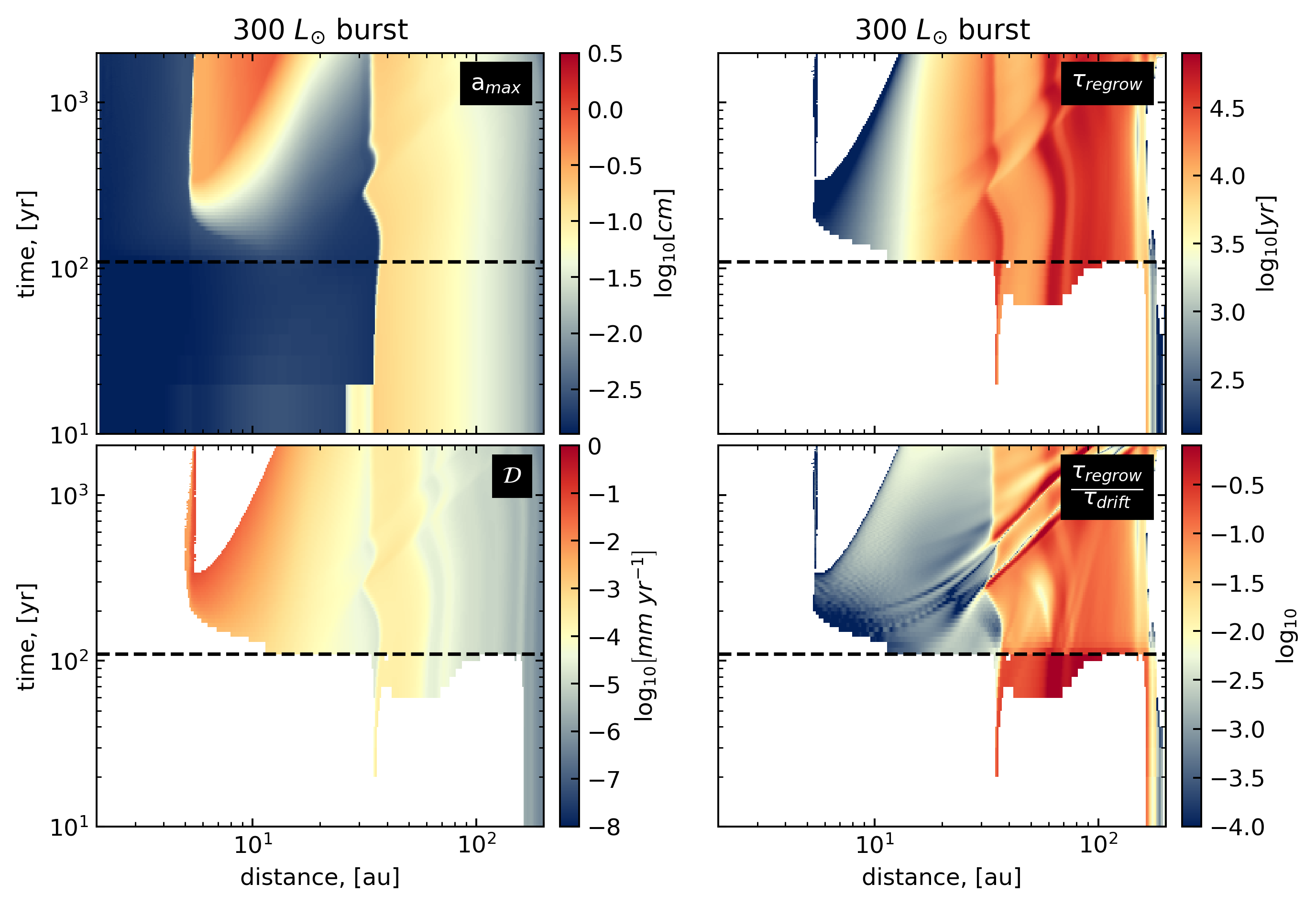} 
\par \end{centering}
\caption{Temporal evolution of the following dust characteristics { in the $300~L_\odot$ model~2}: maximal dust grain size (top-left), dust growth rate (bottom-left), characteristic timescale of dust growth (top-right), and ratio of the characteristic timescale of dust growth to the grown dust drift timescale (bottom-right). 
%The left and right columns correspond to model~2 without and with the burst, respectively. 
Disk regions where dust growth is halted by the fragmentation barrier are filled with white color.  The horizontal black dashed lines indicate the terminal time of the burst.  } 
\label{fig:rates}
\end{figure}

To better understand the effect of the burst on the maximum dust size, we consider in Figure~\ref{fig:rates} 
the characteristic timescale of dust re-growth after the burst $\tau_{\rm regrow}$ and the drift timescale of grown dust on the star $\tau_{\rm drift}$. %in model~2 with and without the burst. 
More specifically, $\tau_{\rm regrow}$ and $\tau_{\rm drift}$ are defined as
\begin{eqnarray}
%    &\tau_{\rm regrow}& = {(a^{\rm pb}_{\rm max} - a^{\rm bst}_{\rm max}) \over {\cal D}},  \\
    &\tau_{\rm regrow}& = {\int_{a_{\rm max}^{\rm bst}}^{a_{\rm max}^{\rm pb}} \dfrac{1}{\cal D} \cdot da} \\
    &\tau_{\rm drift}& = {r\over |u_r-v_r|}.
\end{eqnarray}
Here, $a^{\rm pb}_{\rm max}$ is the preburst value of $a_{\rm max}$ at the corresponding radial location in the disk, $a^{\rm bst}_{\rm max}$ the value of $a_{\rm max}$ during the burst, $u_r$ the radial component of the grown dust velocity, $v_r$ the radial component of the gas velocity, and $r$ the radial distance from the star.  The quantity $\tau_{\rm regrow}$ describes the time that it takes for dust grains to grow to the preburst size after being { affected} by the burst. { A luminosity burst leads to the evaporation of icy mantles, thus causing a decrease in the fragmentation velocity $v_{\rm frag}$  (Eq.~\ref{eq:vfrag}), lowering the fragmentation barrier (Eq.~\ref{afrag}), and triggering more efficient dust fragmentation in the affected regions.}
The white area highlights the disk regions where the dust size has reached its maximum value defined by $a_{\rm frag}$ so that dust growth is halted. The top row shows the time evolution of $a_{\rm max}$ for convenience.

When the burst is triggered, the fragmentation barrier drops because of disk heating by stellar irradiation and the maximum dust size also decreases. After the burst, dust begins to re-grow, but $\tau_{\rm regrow}$  is appreciably longer than the burst duration in the disk regions beyond 10--15~au. Furthermore, Figure~\ref{fig:rates} indicates that the dust drift timescale is longer than the dust growth timescale, meaning that the radial dust drift is not expected to affect the dust spatial distribution on the timescales of several thousand years. All this suggests that the effects of the burst on the dust size distribution can linger for a time interval that is notably longer than the burst duration, opening a possibility for inferring the past luminosity bursts, for instance, using the spectral indices as will be shown below.
%\citep{Pavlyuchenkov2019}.

\subsection{Evolution of the spectral index in the postburst phase}

In this section, we use a simplified model to calculate the radial distribution of the spectral index assuming a local plane-parallel disk geometry and dust temperature that is constant (or weakly changing) in the vertical direction. { Dust settling is implicitly taken into account when calculating the dust volume density in Eq.~(\ref{rho:dust}), which considers that small and grown dust may have generally different vertical scale heights depending on the values of $\alpha_{\rm visc}$ and $\mathrm{St}$. However, this effect is not expected to explicitly influence the spectral index calculations because we consider only face-on disk configurations.} We note that in our model we make no distinction between the gas and dust temperatures, which is justified for the bulk of the disk at the solar metallicity \citep{Vorobyov-lowz2020}. We also note that in the plane of the disk the temperature is computed self-consistently { using the vertically integrated gas pressure and gas density as $T_{\rm mp}=\mu {\cal P}/(\Sigma_{\rm g} {\cal R}$), where $\mu=2.33$ is the mean molecular weight and $\cal R$ is the universal gas constant},  and it can vary significantly both radially and azimuthally throughout the disk. A formal solution of the radial transfer equation in the plane-parallel limit can be written as
\begin{equation}
 I_{\nu}(r,\phi)=B_{\nu}(T_{\rm d})(1-e^{-\tau_{\nu}}),
 \label{eq:modified-black-body}
\end{equation}
where $I_{\nu}$ is the radiation intensity at a given position ($r,\phi$) in the disk, $B_{\nu}$ (erg~s$^{-1}$~cm$^{-2}$~Hz$^{-1}$~sr$^{-1}$) is the Planck function, $T_{\rm d}$ is the dust temperature, and $\tau_{\nu}=\kappa_{\rm \nu,sm}\Sigma_{\rm d, sm}+\kappa_{\rm \nu,gr}\Sigma_{\rm d, gr}$ is the total optical depth of the small and grown dust populations. The frequency dependent absorption opacities $\kappa_{\rm \nu,sm}$ and $\kappa_{\rm \nu,gr}$ (per gramm of dust mass)  for the small and grown dust populations with maximum sizes $a_\ast$ and $a_{\rm max}$, respectively, were taken from our earlier work  \citep{Skliarevskii2021}. We assumed that each dust component has a constant slope of $p=3.5$ for the dust size distribution that extends from $5\times10^{-3}$~$\mu$m to $a_\ast=1.0$~$\mu$m for small dust and from $a_\ast$ to $a_{\rm max}$ for grown dust. The opacities were then calculated using the Mie theory assuming pure silicate grains of spherical shape. We neglect the scattering effects in this study.  We confirmed that our frequency- and size--dependent  opacities agree well with those derived using the `OpacityTool' of \citet{Woitke2016} (see Appendix~\ref{Opacity}). 

\begin{figure*}
\includegraphics[width=\linewidth]{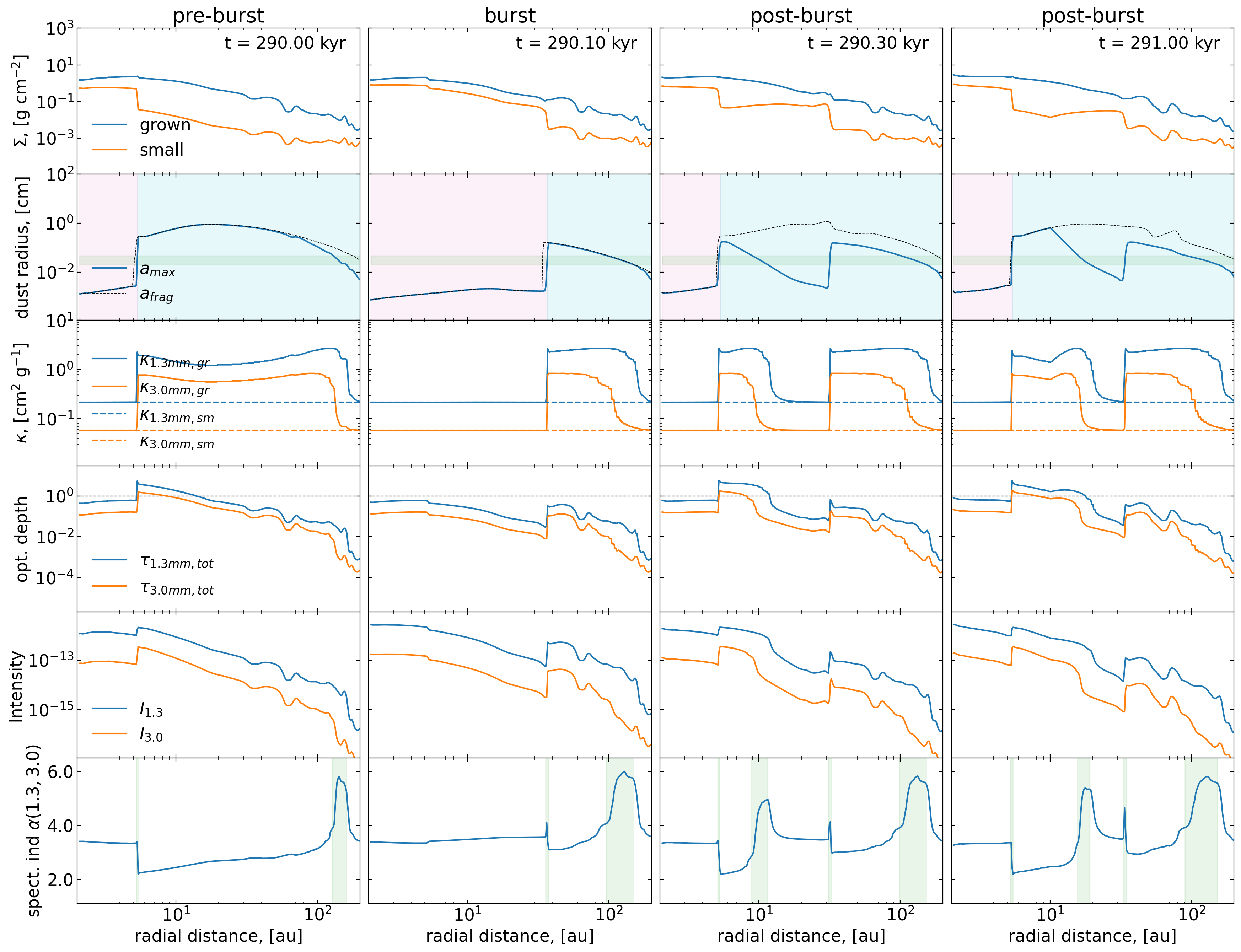}
\caption{Azimuthally averaged disk characteristics before, during, and after the burst. Columns from left to right correspond to the time just before the burst onset (290.0~kyr), just before the end of the burst (290.1~kyr), 200~yrs after the end of the burst (290.3~kyr), and 900~yr after the end of the burst (291~kyr).   Rows from top to bottom show the azimuthally averaged surface densities of small and grown dust, maximum dust size, absorption opacities of small and grown dust at 1.3~mm and 3.0~mm, optical depths at 1.3~mm and 3.0~mm, intensity of dust thermal emission at 1.3 and 3.0~mm (in ergs~cm$^{-2}$~s$^{-1}$~Hz$^{-1}$~St$^{-1}$), and spectral index $\alpha_{1.3-3.0}$.
The pink and blue areas in the second row indicate the disk regions without and with water ice, respectively; the boundary between them is the current position of the water snowline. The black dotted line in the second row shows the fragmentation barrier $a_{\rm frag}$.  The horizontal green strip in the second row is confined by $1.3~\mathrm{mm} / (2 \pi)$ and $3.0~\mathrm{mm} / (2 \pi)$. The crossing of this strip with the $a_{\rm max}$ curve defines the disk regions where the maximum dust size lies in the $1.3~\mathrm{mm} / (2 \pi) < a_{\rm max} < 3.0~\mathrm{mm} / (2 \pi) $    limits. These regions  are also characterized by local peaks or drops in $\alpha(\lambda_1,\lambda_2)$ highlighted by the vertical green strips in the bottom row. The horizontal black dotted line in the forth row shows the optical depth of 1.0 for convenience.}
\label{fig:opac}
\end{figure*}

The spectral index $\alpha(\lambda_1,\lambda_2)$ was calculated using the following equation
\begin{equation}
\label{alpha_spectral}
\alpha(\lambda_1,\lambda_2)=
\dfrac
{\log_{10}{I_{\nu_1}} - \log_{10}{I_{\nu_2}} } 
{\log_{10}{\nu_1} - \log_{10}{\nu_2}},
\end{equation}
where $\lambda_1$ and $\lambda_2$ were set equal to 1.3~mm and 3.0~mm, corresponding to the ALMA Bands 6 and 3, respectively. A large enough difference between the wavelengths also allows us to minimize the  oscillations in $\alpha(\lambda_1,\lambda_2)$, which occur at $\Delta \lambda \lesssim 1.0$~mm \citep{Pavlyuchenkov2019}.

Figure~\ref{fig:opac} demonstrates how the spectral index and the pertinent dust characteristics are changing as the burst develops and decays. The grown dust component dominates the total dust surface density, as can be expected for the chosen slope of the dust size distribution, $p=3.5$. The surface density of small dust exhibits a steep gradient at the position of the water snowline. This feature follows the radial position of the snowline as the burst evolves. At $t=290.3$~kyr (i.e., 200~yr after the burst), two such features can be seen: one corresponding to the current radial location of the snow line and the other reflecting the maximum extent of the snowline at the end of the burst. The evolution of the maximum dust size $a_{\rm max}$ follows a similar pattern. In the preburst and burst phases, $a_{\rm max}$ shows a steep gradient at the water snowline, having smaller/larger values interior/exterior to its current position. This drop is caused by the corresponding drop in the fragmentation barrier $a_{\rm frag}$ (see Eq.~\ref{afrag}).
In the postburst evolution, the snow line retreats to its original location and dust starts to re-grow, but this process is not uniform and proceeds faster in the inner disk regions. As a result, a second steep gradient in $a_{\rm max}$ at a radial distance where the snowline was during the burst can be seen for at least 1~kyr after the burst. We note that $a_{\rm max}$ was limited by the fragmentation barrier before the burst in the disk regions interior $\approx 30-40$~au (see also Fig.~\ref{fig:rates}). Nine hundred years after the burst $a_{\rm max}$ is fragmentation-barrier-limited inside $\approx$~10~au, meaning that the size of dust grains has not yet reached its maximum value set by $a_{\rm frag}$.

\begin{figure}
\includegraphics[width=\linewidth]{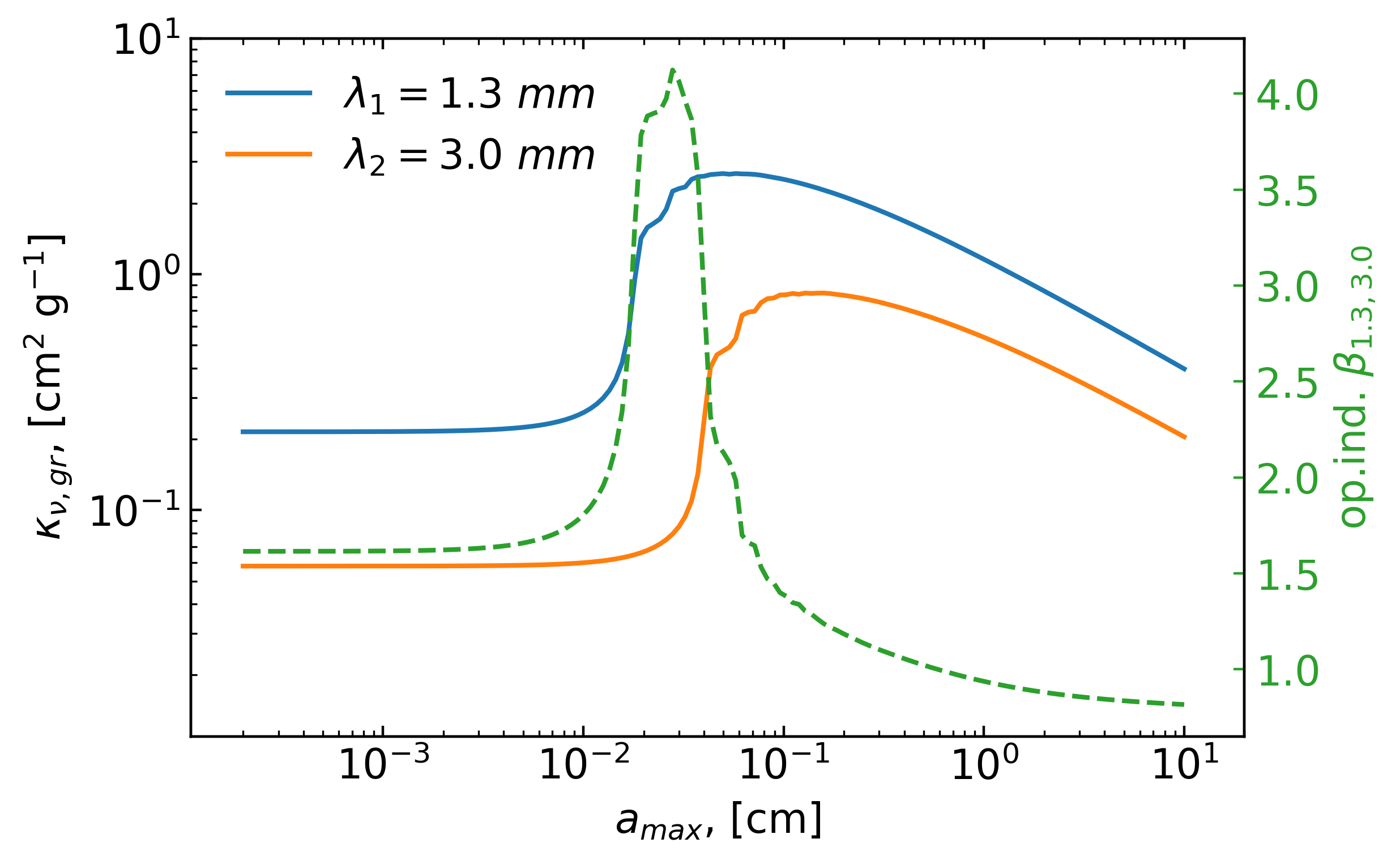}
\caption{Dependence of the absorption opacity  $\kappa_{\rm \nu,gr}$  (blue and orange lines) and the opacity index $\beta$ (dashed green line) on the maximum size of grown dust $a_{\rm max}$. Two wavelengths of $\lambda_1=1.3$~mm and $\lambda_2$=3.0~mm are considered, which correspond to the ALMA Band 6 and 3, respectively. }
\label{fig:cliff}
\end{figure}

Figure~\ref{fig:opac} demonstrates that the opacities of small and grown dust do not differ from each other in the disk regions with $a_{\rm max}\la 10^{-2}$~mm, which can be expected for dust absorption at millimeter wavelengths. However, in the disk regions where $a_{\rm max}\ga 0.1$~mm, the total opacity is dominated by grown dust. There is a steep gradient in the grown dust opacity at the water snowline, which is a manifestation of the phenomenon known as the opacity cliff, namely, a sharp drop in the dust opacity as the maximum size of dust grains falls below $ \lambda / 2\pi$ (see Fig.~\ref{fig:cliff}). As the water snow line advances outward during the burst, the sharp gradient in $\kappa_{\nu,\rm gr}$  moves concurrently. When the snowline recedes after the burst, this feature moves back to the preburst location. However, because of a non-uniform dust re-growth, another jump in $\kappa_{\nu,\rm gr}$ can be seen at $\approx 35$~au for at least a thousand years. We also note that a drop in $a_{\rm max}$ in the outer disk regions beyond 100~au causes the grown and small dust opacities to converge quickly to a similar value.

The strong dependence of grown dust opacity on $a_{\rm max}$ has a notable effect on the radial profile of the total optical depth at both considered wavelengths. In particular, $\tau_\nu$
decreases during the burst so that the disk becomes optically thin. We note that this may not be the case, in particular, for the innermost disk regions if the $\alpha_{\rm visc}$-value is globally much lower than 0.01 \citep{Banzatti2015}. After the burst, part of the disk becomes optically thick again because of dust re-growth. This suggests that the disk mass estimates in the burst phase are likely to be more accurate than in the preburst/postburst stages. 
This can make FUors an important laboratory, given that disk mass estimates are not always consistent and suffer a lot from incorrect assumptions on dust properties, including the optical depth \citep[e.g.,][]{Ballering2019}.
Furthermore, the total optical depth exhibits steep gradients at the position of the water snowline during and after the burst, similar to what has been discussed above in the context of grown dust opacities. 
A similar jump in the optical depth was also reported in, for example, \citet{Banzatti2015} for a simplified steady-state model of the $\alpha_{\rm visc}=0.01$ disk. The radial profiles of intensity of the dust emission at 1.3 and 3.0~mm follow closely in shape those of the corresponding optical depths, except for the optically thick disk regions where the sharp jumps are smoothed.

Finally, we consider the radial profiles of the spectral index $\alpha(1.3,3.0)$ shown in the bottom row of Figure~\ref{fig:opac}. The position of the water snowline before the burst is characterized by a sharp drop in the spectral index to a value of $\approx 2.0$, reflecting the transition from optically thin to optically thick disk regions. During the burst ($t=290.1$~kyr), this feature shifts to a larger distance (together with the snowline) but the amplitude of the drop in $\alpha(1.3,3.0)$ decreases. Concurrently, the spectral index is rising, reflecting the overall decrease in the optical depth during the burst. After the burst, the sharp drop in $\alpha(1.3:3.0)$ moves back to the original location, while the corresponding feature at $\approx35$~au transforms into a very narrow peak. In addition, a broad local peak in the spectral index appears at 10~au and moves slowly outward. This peak is caused by spatially non-uniform dust re-growth to the original preburst size. At all times, there is another broad peak in  $\alpha(1.3,3.0)$ localized beyond 100~au, which may be hard to detect due to the overall drop of dust density in these outer disk regions.

The occurrence of a local peak in the spectral index is illustrated in Figure~\ref{fig:cliff} showing the dependence of the absorption opacity $\kappa_{\rm \nu,gr}$ on the maximum size of grown dust $a_{\rm max}$.
The absorption coefficient experiences a sharp drop at a certain value of $a_{\rm max}$, which is distinct for different wavelengths.  This drop may cause apparent opacity gaps in protoplanetary disks, which do not correspond to any physical gaps and shift with the observational wavelength \citep{Akimkin2020}.  As the result, the opacity index $\beta$ calculated as
\begin{equation}
\beta(\lambda_1,\lambda_2)=
\dfrac
{\log_{10}{\kappa_{\rm \nu_1,gr}} - \log_{10}{\kappa_{\rm \nu_2,gr}} } 
{\log_{10}{\nu_1} - \log_{10}{\nu_2}}
\end{equation}
exhibits a local peak in the vicinity of the opacity cliff.
In the optically thin medium and in the Rayleigh-Jeans limit the spectral and opacity indices are related to each other as $\alpha=2+\beta$, meaning the presence of the corresponding peak in the spectral index as well.

\citet{Cieza2016} have reported the presence of a broad peak in the radial distribution of spectral indices in V883~Ori, an FU-Orionis-type object with the unknown time of eruption. These authors interpreted the broad peak in $\alpha(\lambda_1=1.29~\mathrm{mm},\lambda_2=1.38~\mathrm{mm})$ as the result of pile-up of small dust at the water snowline \citep[see also][]{Banzatti2015}. \citet{Schoonenberg2017} noted that the pile-up timescale is much longer than the burst duration of 10--100 yr and proposed that destruction of large dust grains (glued together by ice mantles) in the vicinity of the snowline that moves outward with the burst can account for the observed peak in the spectral index. 

Our work is not aimed at explaining the spectral index distribution in V883~Ori. Nevertheless, the position of the observed peak in V883~Ori at $\approx~42$~au agrees with the position of the sharp peak at $\approx 35$~au in model~2 (second column in Fig.~\ref{fig:opac}), but the shapes of the observed and model features are not consistent with each other -- the observed peak is much broader. Taking the model caveats into account in our future works (see Sect.~\ref{caveats}) may help to achieve a better agreement. If we turn to the postburst evolution, however, a broad peak in $\alpha(1.3,3.0)$ that forms  at $\approx 10$~au (third and fourth columns in Fig.~\ref{fig:opac}) agrees much better in shape with the feature detected in V883~Ori. This peak then moves outward with time.
In our toy model of the outburst, the stellar luminosity  abruptly returns to a preburst value after 100~yr. This is certainly an oversimplification. After quickly achieving a peak value,  the light curve may slowly decline to a preburst state. This may initiate changes in the disk temperature and the corresponding radial distribution of dust sizes, leading to the formation of a broad peak in $\alpha(\lambda_1,\lambda_2)$ that gradually moves outward, similar to what is found in our model. Combined with the unknown onset date of V883~Ori outburst (there are indications that the object was in outburst already in 1890, \citet{Pickering1890}), this may point to a stronger burst amplitude in the past  followed by a gradual decline that likely exceeds several centuries in duration. Indeed, \citet{Sandell2001} give a value of $400~L_\odot$, while  more recent estimates suggest a lower value of $\approx 210~L_\odot$ \citep{Connelley2018}.
We plan to check these conjectures in a follow-up study.

Interestingly, drops and local peaks in $\alpha(\lambda_1=1.3~\mathrm{mm},\lambda_2=3.0~\mathrm{mm})$  are located in the disk regions where the maximum dust size falls in the $ 1.3~\mathrm{mm} / (2 \pi) < a_{\rm max} < 3.0~\mathrm{mm} / (2 \pi) $  limits (see the green shaded horizontal and vertical strips in Fig.~\ref{fig:opac}). 
%In general, the peak positions in $\alpha(\lambda_1,\lambda_2)$ are defined by the crossing of a horizontal bar having  boundaries $\lambda_1/2 \pi$ and $\lambda_2/2 \pi$  and the curve showing the radial distribution of $a_{\rm max}$ (see the second and bottom rows in Fig.~\ref{}).
It was previously suggested that the local peaks in the spectral index can be used to infer the maxim size of dust grains \citep[e.g.,][]{Pavlyuchenkov2019}. It appears that the local peaks in the spectral index that naturally develop in the disks of FUors can also be used for this purpose.

\section{Parameter space study}
\label{parstudy}
In this section, we perform a parameter space study to determine how the evolution of $a_{\rm max}$ and $\alpha(\lambda_1,\lambda_2)$ can be affected by the burst magnitude and disk morphological structure. We note that all previous results were obtained for a burst magnitude of $300~L_\odot$. Here we also consider a case with a burst magnitude of $100~L_\odot$, which is further referred to as model~3. The burst in this model is initiated at $t=290$~kyr.  In addition, we introduce a new model (referred to as model~4) in which we relax the constant $\alpha_{\rm visc}$-parameter assumption. In this model, the $\alpha_{\rm visc}$-value is defined according to the layered disk model of \citet{Armitage2001} further refined by \citet{Bae2014}. More specifically, the adaptive $\alpha_{\rm visc}$-parameter is calculated as
\begin{equation}
\alpha_{\rm visc}={\Sigma_{\rm g,act} \,\, \alpha_{\rm max} + \Sigma_{\rm g,dead} \,\, \alpha_{\rm dead} \over \Sigma_{\rm g,act} + \Sigma_{\rm g,dead}}.
\label{alphaV}
\end{equation}
The MRI-active vertical column of the disk and the corresponding $\alpha_{\rm visc}$-value are set equal to $\Sigma_{\rm g,act}=100$~g~cm$^{-2}$  and $\alpha_{\rm max}=0.01$, respectively. 
The $\alpha_{\rm visc}$-value of the MRI-dead column of the disk $\Sigma_{\rm g,dead}$ is set to $\alpha_{\rm dead}=10^{-5}$. The MRI-dead vertical column of the disk $\Sigma_{\rm g,dead}$ is set equal to $\Sigma_{\rm g}/2 - \Sigma_{\rm g, act}$ if the latter expression is positive and to zero otherwise.

Equation~(\ref{alphaV}) indicates that the intermediate and outer disk regions with gas surface densities $\le 200$~g~cm$^{-2}$ are MRI-active, while in the innermost disk regions where $\Sigma_{\rm g}$ can greatly exceed 200~g~cm$^{-2}$ the $\alpha_{\rm visc}$-value effectively reduces to $\simeq10^{-5}$. These latter regions of reduced viscous mass transport constitute a `dead' zone where matter accumulates after being transported by viscous and/or gravitational torques from the disk outer regions. 
The key parameters of all considered models are listed in Table~\ref{table:1} for convenience.

\begin{table}
\center
\caption{\label{table:1}Model parameters}
\begin{tabular}{ccccc}
\hline 
\hline 
Model & Peak lum. & $\alpha_{\rm visc}$ &  disk age & disk mass   \tabularnewline
number & [$L_\odot$] &  &  [kyr] & ($<100$~au) [$M_\odot$]  \tabularnewline
\hline 
1 &  300 & 0.01 &  90 & 0.26 \tabularnewline
2 &  300 & 0.01 &  290 & 0.17 \tabularnewline
3 &  100 & 0.01 &  290 & 0.26 \tabularnewline
4 &  300 & adaptive &  340 & 0.36 \tabularnewline
%4 &  300 & 0.01 &  390 & 0.5:5 \tabularnewline
\hline 
\end{tabular}
\end{table}

\begin{figure}
\includegraphics[width=\linewidth]{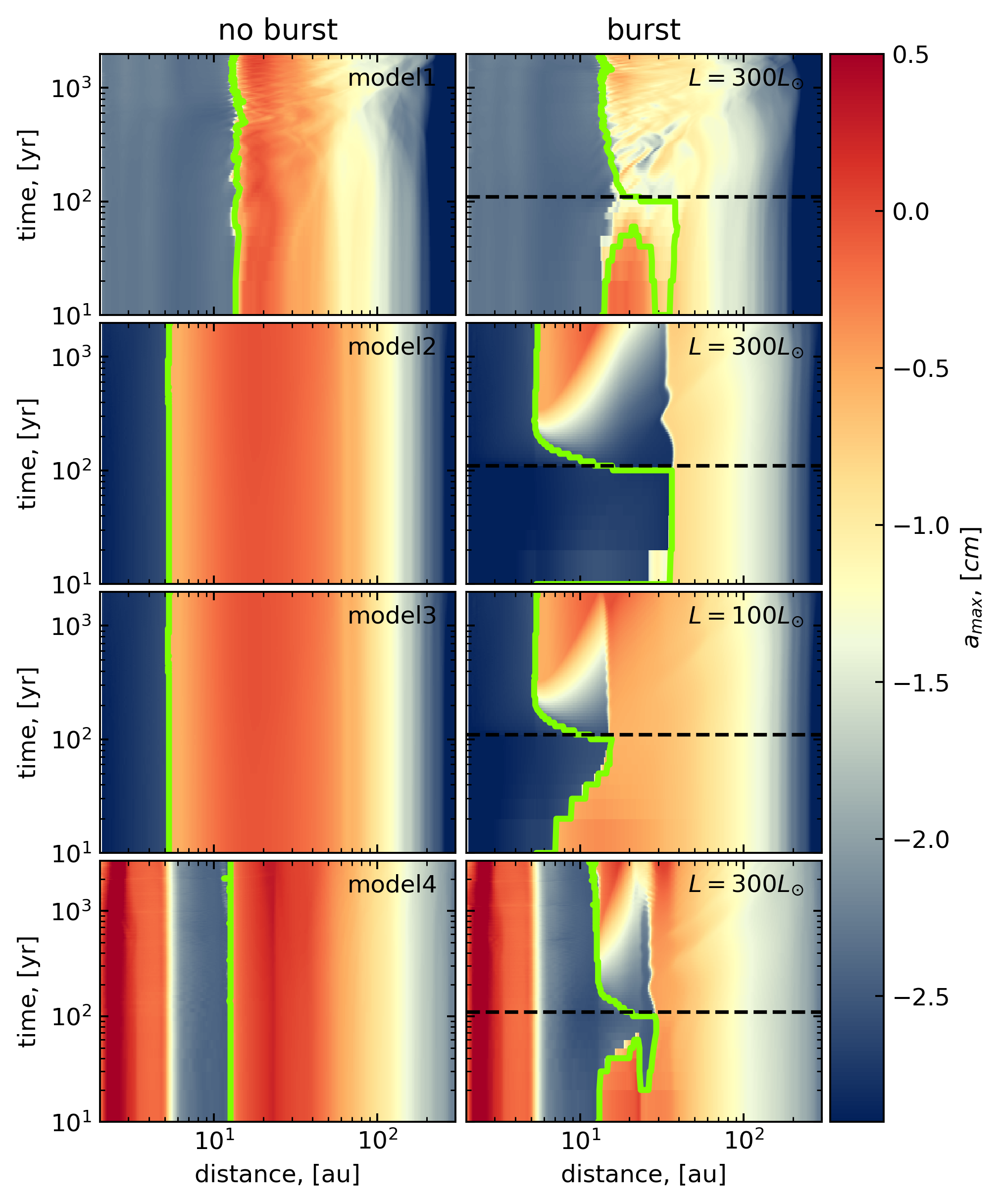}
\caption{Space-time diagrams showing the time evolution of the azimuthally averaged maximum size of dust grains $\overline{a}_{\rm max}$. Rows from top to bottom  correspond to distinct models, while the left and right columns show the cases without and with the burst. The green curves outline the position of the water snowline. The horizontal dashed line in the left column indicates the end of the burst.}
\label{fig:dustsize}
\end{figure}

Figure~\ref{fig:dustsize} presents the space-time diagrams showing the time evolution of the azimuthally averaged maximum dust size $\overline{a}_{\rm max}$ in all four considered model. The right column shows the effect of the burst, while the left column provides the corresponding data in a quiescent disk for comparison. The effect of the burst on the time evolution of $\overline{a}_{\rm max}$ is most notable and expressed in model~2, which corresponds to an evolved nearly-axisymmetric disk with a maximum considered burst magnitude of $300~L_\odot$. The burst in model~1 has the least pronounced effect. The disk in this model is characterized by a complex non-axisymmetric structure (see Figs.~\ref{fig:init} and \ref{fig:endburst}). In particular, the H$_2$O snowline cannot be described by a simple circle and the azimuthal averaging washes out the local variations in $a_{\rm max}$ caused by the burst. This means that young gravitationally unstable disks may not be the best candidates for using $a_{\rm max}$ as a tracer of the past burst because of their complex spatial morphology. A higher angular resolution would also be required to resolve the fine disk structures such as spiral arms.   The burst with a lower peak luminosity in model~3 has a weaker impact on $\overline{a}_{\rm max}$, but the effect is still noticeable for about a thousand years.

Model~4 with an adaptive $\alpha_{\rm visc}$-parameter has a spatial distribution of $\overline{a}_{\rm max}$ in the quiescent disk that notably differs from other considered cases. In particular, the maximum size of dust grains interior of the H$_2$O snowline strongly increases at small radii, reaching values that are comparable to or even greater than those immediately beyond the snowline. { More specifically, $\overline{a}_{\rm max}$ in the inner 5 au of the constant $\alpha_{\rm visc}$ model~2 is 10--60 micron, while in the adaptive $\alpha_{\rm visc}$ model~4  its value lies  in the 7--50 mm limits. Beyond the snow line, $\overline{a}_{\rm max}$ in model~4 does not exceed 10~mm. }
This is the effect of decreasing $\alpha_{\rm visc}$ in the dead zone at several astronomical units and, as a consequence, increasing $a_{\rm frag}$ (see Eq.~\ref{afrag}). The water snowline is located at a larger distance in this model because of warmer inner parts of the disk occupied by the dead zone. 
Interestingly, the burst does not affect the distribution of $\overline{a}_{\rm max}$ inside the water snowline because the temperature is set there mostly by viscous and compressional heating, and not by stellar irradiation \citep{2018VorobyovAkimkin}. 
On the other hand, the burst has a notable effect on the distribution of $\overline{a}_{\rm max}$  outside of the snowline where stellar irradiation is the dominant source of energy. The effect of the burst on $\overline{a}_{\rm max}$ can last for more than a thousand years after the burst has ended, yet it is restricted to a narrower radial annulus of the disk as compared to model~2 because of a steeper temperature gradient in the adaptive $\alpha$-model (the inner disk regions occupied by the dead zone are optically thicker and hence warmer in this model).

\begin{figure}
\includegraphics[width=\linewidth]{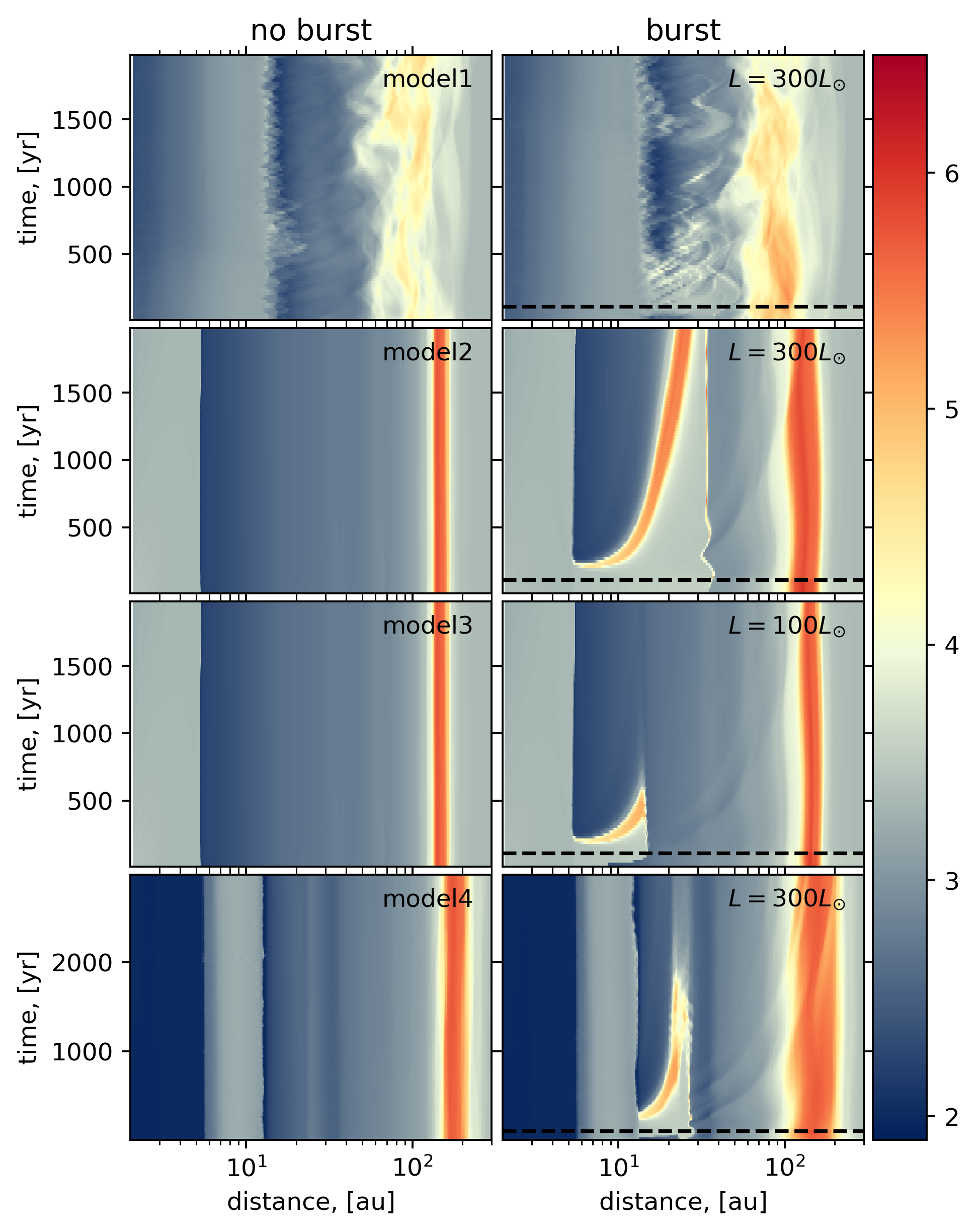}
\caption{Space-time diagrams showing the time evolution of the azimuthally averaged spectral index $\overline{\alpha}(\lambda_1,\lambda_2)$ with $\lambda_1=1.3$~mm and $\lambda_2=3.0$~mm. Each row corresponds to the indicated model, while the left and right columns show the cases without and with the burst, respectively.  The horizontal dashed line in the left column indicates the end of the burst.}
\label{fig:specind}
\end{figure}

Figure~\ref{fig:specind} presents the space-time diagrams of the azimuthally averaged spectral index $\overline{\alpha}(\lambda_1,\lambda_2)$ with $\lambda_1$ and $\lambda_2$ set equal to 1.3 and 3.0~mm, respectively. The right column presents the time evolution of $\overline{\alpha}(1.3,3.0)$ in the burst models, while the left column shows the corresponding data in the quiescent models for comparison. Model~1 with a young non-axisymmetric disk shows a rather complex distribution of the spectral index, reflecting the complexity of the disk spiral morphology (see Figs.~\ref{fig:init} and \ref{fig:endburst}). There are no clear signatures of the past burst in the azimuthally averaged data, meaning that the two-dimensional distribution of $\alpha(1.3,3.0)$ needs to be analysed, which is more complex and is planned for the future work. The effect of the burst on the radial distribution of the spectral index is most pronounced in model~2, which features a strong luminosity burst and an evolved nearly-axisymmetric disk. A broad inner peak in $\overline{\alpha}(1.3,3.0)$ can be seen moving outward from a radial distance of $\approx 10$~au just after the end of the burst to $\approx 25$~au at  two thousands of years after the burst. We have not extended our simulations beyond 2~kyr but by extrapolation this feature may last for up to 10~kyr. Models~3 and 4 also feature such a peak in $\overline{\alpha}(1.3,3.0)$ but it persists for a shorter time and disappears after just 1-2~kyr. A lower magnitude burst has a weaker effect on the dust maximum size, so that dust can re-grow to its original distribution faster. 

{ Observations of silicate features in FUors indicate that outbursts are more common in young systems that are still embedded in parental contracting cores than in more evolved optically revealed counterparts \citep{Quanz2007}. Numerical simulations confirm this trend \citep{VorobyovBasu2015,Vorobyov-mhd2020}. However, the presented analysis of the spectral index can still be applicable to young disks if they are not prone to the development of strong gravitational instability. This may be the case when the disks are warmed up by external irradiation or have insufficient mass for gravitational instability to develop or stabilized by strong magnetic fields. V883 Ori may be an example of such a system showing the silicate features in absorption (young system) but having an apparently axisymmetric disk.}

In the adaptive $\alpha_{\rm visc}$-value model~4 the radial distribution of $\overline{\alpha}(1.3,3.0)$ is more complex. The innermost 5~au are optically thick because of accumulation of dust in the dead zone. The spectral index there approaches 2.0. In the 5-10~au region between the outer edge of the dead zone and the water snowline  the spectral index has a local maximum with $\overline{\alpha}(1.3,3.0)\approx 3.2$. This is the disk region where the size of dust grains and the corresponding optical depth decrease, making the disk optically thin. Beyond the snowline the size of dust grains and the corresponding optical depth increase and the spectral index drops again.
In this model, the burst creates another strong peak in $\overline{\alpha}(1.3,3.0)$ located between 10 and 25 au.
The effect of the burst is limited to a narrower radial annulus of the disk, making the burst-triggered peak live shorter than in model~2 with the same burst magnitude.

We argue that the presence of a broad peak in the spectral index at a few tens of au in the disk of a non-burst star may signalize a strong luminosity burst in the recent past. This peak also changes position with time, but this change may be difficult to detect over the human lifetime. Similar peaks in $\overline{\alpha}(\lambda_1,\lambda_2)$ can be present in the disk of quiescent stars \citep[e.g.][]{Banzatti2015}, but they are expected to be localized either within the inner 10~au from the star where normally the water snowline and the dead zone are localized or outside 100~au where they may be difficult to detect because of low dust densities.

\section{Model caveats and further steps}
\label{caveats}
In this section, we discuss several model caveats that need to be addressed in future studies. 

{\it Burst light curve}.  The burst was modeled using the simplest approach in which the stellar luminosity was increased to a fixed peak value of either 100~$L_\odot$ or 300~$L_\odot$. The duration of the burst was fixed in all models at 100~yr. The burst ignition was also limited to the inner 2~au, which is the size of our sink cell. In reality, FUors exhibit a variety of light curves with distinct burst magnitudes, duration, and shape \citep{Audard2014}. The extent of the inner disk region involved in the burst triggered by the magnetorotational instability may also exceed 2~au but will likely stay within 10~au \citep{Vorobyov-mhd2020}, thus not affecting our region of interest. In follow-up studies, we plan to explore the effects of a self-consistent luminosity outburst caused by the magnetorotational instability in the innermost disk regions based on the MHD accretion burst model developed in \citet{Vorobyov-mhd2020}. It would also be interesting to consider the effects of recurrent outbursts with the quiescent period between the bursts on the order of hundred or thousand years, as suggested by the clustered burst model of \citet{VorobyovBasu2015} or the analysis of jet knot spacing in CARMA~7 \citep{Vorobyov-jets2018}.

{\it Viscous $\alpha_{\rm visc}$-parameter}. The $\alpha_{\rm visc}$-parameter was fixed to a constant value of 0.01 in most models. We have also considered the effect of an adaptive viscous parameter, in which case the $\alpha_{\rm visc}$-value can drop substantially in the inner dead zone (see Eq.~\ref{alphaV}). However, protoplanetary disks may feature globally lower values of $\alpha$  down to $10^{-3}$ or even $10^{-4}$ \citep[e.g.,][]{Flaherty2020}. As was shown by \citet{Banzatti2015}, the radial distributions of the spectral index in the millimeter dust emission are different throughout the disk extent when globally lower values of the $\alpha_{\rm visc}$-parameter are considered. It is therefore important to consider in the follow-up study the cases with globally lower turbulent viscosity.

{\it Dust growth and destruction}. In the FEOSAD code, the dust growth is calculated following the monodisperse model of \citet{Stepinski1997}. We do not consider sophisticated dust growth effects, such as preferential recondensation of water ice on dust nuclei after the burst \citep{Hubbard2017}. The growth of dust is stopped when $a_{\rm max}$ reaches the fragmentation barrier $a_{\rm frag}$. However, the destruction of dust grains when $a_{\rm max}$ drops below  $a_{\rm frag}$ (e.g., when the burst heats the disk and raises the temperature) is treated in a simplistic manner by setting $a_{\rm max}=a_{\rm frag}$. In the future studies, a more sophisticated model of dust growth and destruction has to be considered, which would take into account mutual collisions between different dust fractions as was done in, for example, \citet{Akimkin2020}, with the
ultimate goal of solving the Smoluchowski equation for dust
coagulation \citep[e.g.,][]{2019Drazkowska}.

{\it Dust composition}.
As was demonstrated by \citet{Schoonenberg2017}, the shape of the peak in the spectral index may depend on dust composition, shape, and porosity. In the current work, we have considered a simple case with silicate dust grains of spherical shape. The effects of more complex dust compositions taking, for example, carbonaceous materials, ice mantles, and porosity into account are also worth studying in the future.  

{\it Dust fragmentation velocity}. The dust fragmentation velocity may vary by a factor of several. For instance, \citet{Okuzumi2016} have used 0.5~m~s$^{-1}$ and 5.0~m~s$^{-1}$ for the bare and water-ice-covered grains. The maximum size of dust grains is sensitive to the fragmentation velocity. The increase/decrease in $a_{\rm max}$ due to variations in $v_{\rm frag}$ may affect the timescale of dust regrowth after the burst and dust drift efficiency, and hence should be explored in follow-up studies. We also note that recent laboratory experiments suggest that bare silicate grains may be at least as sticky as water-coated grains \citep[e.g.,][]{Gundlach2018,Musiolik2021}. This finding may be supported by observations and modeling of the dust spectral index in FU~Ori \citep{LiuVorobyov2019}. We plan to explore this possibility in future studies.   

{\it Dust turbulent diffusion.} {   In this work we neglected the effects of turbulent diffusion of dust.  To estimate its importance, we calculated the viscous time scale $t_{\rm visc} = r^2/\nu$, which should be a good proxy for the diffusion timescale given that the Stokes numbers in our models are in general lower than 0.1. The resulting $t_{\rm visc}$ increase from a few~$\times 10^3$~yr at $r=2$~au to $\approx 10^6$~yr at $r=100$~au, with typical values on the order of $10^5$~yr in the region of interest $r=10-50$~au. Clearly, the viscous timescales are much longer than both the lifetime of the burst and the maximum age of the peak in the spectral index. Therefore, dust turbulent diffusion is not expected to alter our results. }

{\it Disk synthetic images}. The observability of the peaks in the spectral index may be an issue, considering that most FUors are located at distances of several hundred to a thousand of parsecs. It is therefore important to check our predictions with radiative transfer simulations combined with CASA postprocessing to derive the disk synthetic images and spectral indices in dust continuum.  These radiative transfer simulations should also include the effects of scattering and vertical temperature gradients, both neglected in the current work \citep[see, e.g.,][]{Liu2019,Sierra2020}. In particular, scattering becomes important in optically thick disk regions, but can tentatively be ignored in the current work because our predictions are 
mostly for optically thin regions.

\section{Conclusions}
\label{conclude}
We studied the response of the gas-dust disk to a luminosity burst that is typical of FU Orionis-type objects. For this purpose we used the FEOSAD numerical hydrodynamics code \citep{2018VorobyovAkimkin,Molyarova2021}, which features the co-evolution of gas, dust, and volatiles in a protoplanetary disk in the thin-disk limit. Dust growth and back reaction on gas were also considered. The burst was modeled by artificially increasing the luminosity of the central star to values in the 100--300~$L_\odot$ range, with the burst duration set equal to 100~years. Two disk configurations were considered: a young gravitationally unstable disk with a well-defined spiral pattern and a more evolved axisymmetric and gravitationally stable disk. The particular emphasis was placed on the evolution of the spectral indices in the millimeter dust emission during and after the burst. Our main conclusions can be summarized as follows. 

-- The burst heats the disk and shifts the water snowline outward, thus expanding the disk region with a lower maximum dust size interior to the snowline position.
%resulting in the drop of the maximum dust size interior to the snowline position. 
In the evolved disk, the shape of the snowline is near circular, while in the young gravitationally unstable disk it can have a complex non-circular geometry \citep[see also][]{Molyarova2021}.
After the burst has ended, the water snowline shifts quickly back to its preburst location and the process of dust regrowth ensues. However, this process is radially non-uniform with the dust sizes in the inner regions reaching the preburst values faster than in the outer regions. 

-- The radially non-uniform renewed growth of dust after the burst creates a broad peak in the spectral index $\alpha(\lambda_1=1.3~\mathrm{mm},\lambda_2=3.0~\mathrm{mm})$ at $\approx 10$~au, which gradually moves outward as the disk evolves. This peak is present in the disk for up to several thousand years after the burst has ended, depending on the maximum luminosity during the burst and radial variations in the viscous $\alpha_{\rm visc}$-parameter. This feature is best expressed in evolved axisymmetric disks, but may not be easily detected in young non-axisymmetric disks because of their complex spatial morphology dominated by spiral arms. { Nevertheless, young disks may still be good candidates to search for the clear peak in the spectral index if gravitational instability is suppressed owing to insufficient disk mass, high external irradiation, or strong magnetic fields.}

-- Local peaks in $\alpha(\lambda_1,\lambda_2)$ that develop in disks of FUors can be used to infer dust maximum sizes, as was already demonstrated for quiescent disks in \citet{Pavlyuchenkov2019}.

-- The presence of a local broad peak in the spectral index was detected in V883~Ori \citep{Cieza2016}, an FU Orionis-type star with the unknown outburst date. Although our work was not focused on reproducing the V883~Ori features, we note a general agreement in the shape and, to a lesser degree, in the position of the model and observed peaks. Our modeling also revealed the formation of other features in the radial distribution of the spectral index in our burst models (such as sharp drops and peaks inside 10~au and beyond 100~au), but their shapes and positions do not agree with those of the observed spectral index peak in V883~Ori.  

{-- We confirmed our earlier conclusion \citep{2018VorobyovAkimkin} that dust does not accumulate appreciably in the spiral arms of an MRI-active disk with $\alpha_{\rm visc}=10^{-2}$.  This is caused by a decrease of the Stokes number below 0.01 within the arms, thus making the dust flow effectively coupled to that of gas (see Appendix~\ref{coupling}).  The efficiency of dust accumulation in the spiral arms may increase in low-turbulence disks where the maximum size of dust grains and the corresponding Stokes number can be larger.
}

As a byproduct, we found that the optical depth at 1.3~mm and 3.0~mm decreases during the burst so that the disk can become optically thin even if it was partly optically thick before the burst. This can make FUors an important laboratory for disk mass estimates, given that these estimates are not always consistent and suffer from incorrect assumptions on dust properties, including the optical depth.

\section*{Acknowledgements} We are thankful to the anonymous referee for constructive comments that helped to improve the manuscript.
E.I.V. and A.S. were supported by the Russian Fund for Basic Research, Russian-Taiwanese project 19-52-52011.  S.-Y.L., H.B.L. and M.T. are supported by MoST of Taiwan 108-2923-M-001-006-MY3 for the Taiwanese-Russian collaboration project.
VA and TM are grateful to the Foundation for the Advancement of Theoretical Physics and Mathematics ’BASIS’ for financial support (20-1-2-20-1). H.B.L. is supported by the Ministry of Science and Technology (MoST) of Taiwan (Grant Nos. 108-2112-M-001-002-MY3).
\'A. K\'osp\'al acknowledges funding from the European Research Council (ERC) under the European Union's Horizon 2020 research and innovation programme under grant agreement No 716155 (SACCRED).
M.T. is supported by the Ministry of Science and Technology (MoST) of Taiwan (grant No. 106-2119-M-001-026-MY3, 109-2112-M-001-019, 110-2112-M-001-044). The simulations were performed on the Vienna Scientific Cluster (VSC-3 and VSC-4).

\begin{appendix}

\section{Toomre parameter and disk stability}
\label{Qparam}

\begin{figure}
\includegraphics[width=\linewidth]{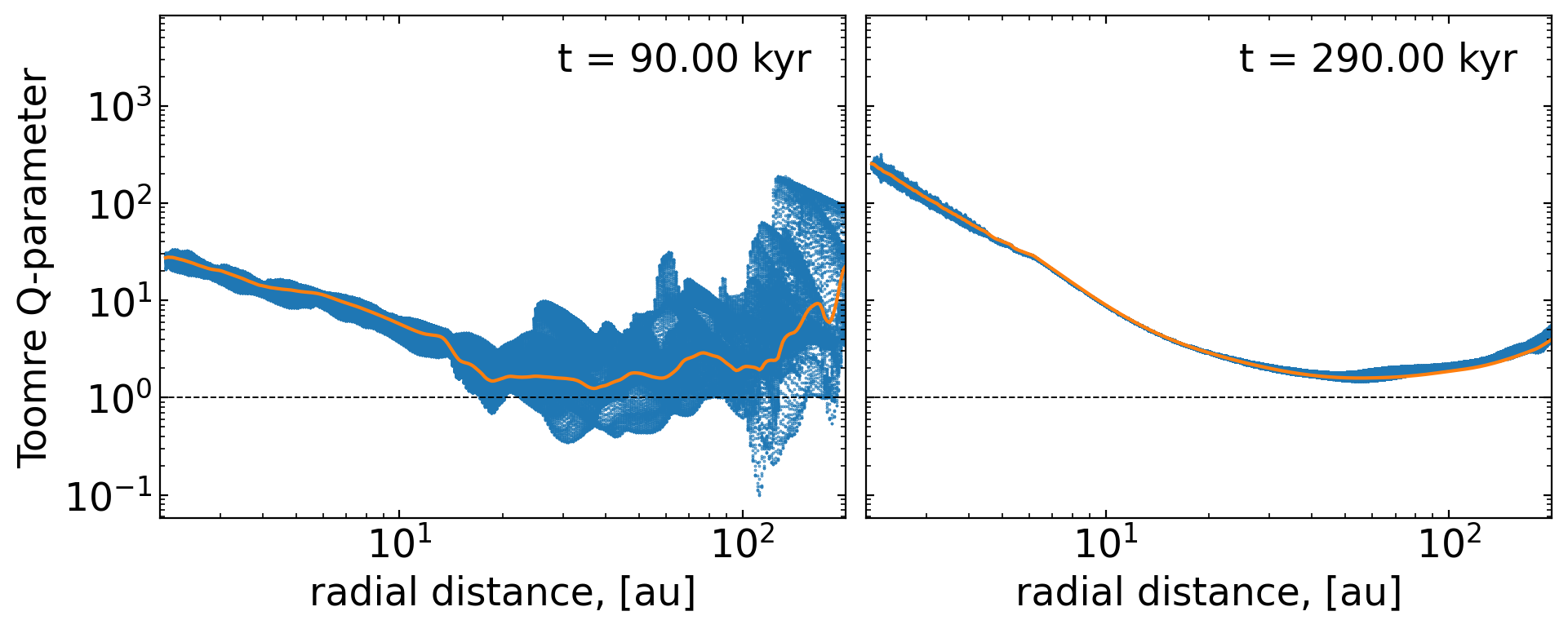}
\caption{Distribution of the Toomre $Q$-parameter in models 1 and 2. The horizontal dotted line shows the critical value for gravitational instability. The orange curves present the azimuthally averaged $Q$-parameter.}
\label{fig:Qparam}
\end{figure}

Here we compare the Toomre $Q$-parameter in young and evolved disks of models~1 and 2, respectively. The $Q$-parameter is often invoked to access gravitational instability in the disk \citep{Toomre1964}.   We define the $Q$-parameter so as to take the contribution of the dust disk into account
\begin{equation}
Q={c_{\rm d} \Omega \over \pi G (\Sigma_{\rm g}+\Sigma_{\rm d,tot})},
\label{ToomreQ}
\end{equation}
where $\Sigma_{\rm d,tot}$ is the total dust surface density, $\Omega$ the local angular velocity,
$c_{\rm d}=c_{\rm s}/\sqrt{1+\zeta_{d2g}}$ the modified sound speed 
in the presence of dust, and $\zeta_{d2g}=\Sigma_{\rm d,tot}/\Sigma_{\rm g}$ the total dust-to-gas ratio. 

Figure~\ref{fig:Qparam} presents the radial distribution of the $Q$-parameter in model~1 (left panel) and model~2 (right panel).
More specifically, the blue dots show variations in the $Q$-values along the azimuth at a fixed radial distance from the star. The orange curves present the azimuthally averaged $Q$-parameter, which is calculated from the corresponding azimuthally averaged surface densities and sound speeds.
The classic stability analysis of \citet{Toomre1964} states that the disk is unstable when $Q$ drops below unity at least in part of the disk. Clearly, the $Q$-values in model 1 drop below 1.0 beyond 15~au, whereas in model~2 this is not the case. The reason for disk stabilization is a gradual increase in the stellar mass (0.34~$M_\odot$ at $t=90$~kyr and 0.46~$M_\odot$ at $t=290$~kyr) and decrease in the surface density (partly through accretion on the star and partly through viscous spreading). The disk also becomes colder with time, which favors the instability, but the effect of disk cooling is overcome by the stellar mass and surface density.

\section{Frequency- and dust-size-dependent opacities}
\label{Opacity}

\begin{figure}
\includegraphics[width=\linewidth]{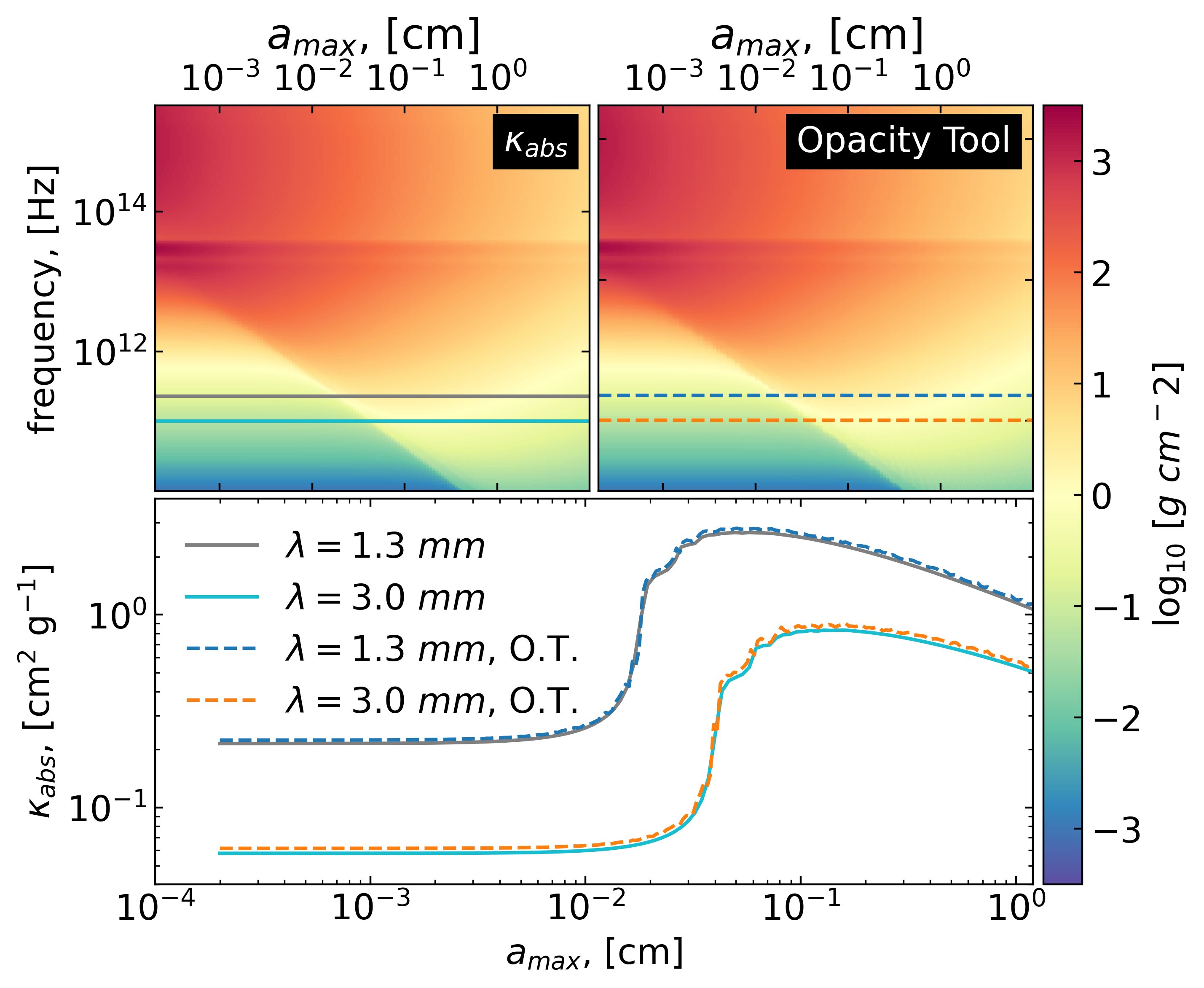}
\caption{Frequency-dependent dust absorption opacities as a function of dust maximum size. Upper-left panel -- data taken from \citet{Skliarevskii2021}; upper-right panel -- data taken from \citet{Woitke2016}. A comparison between the two data sets at 1.3~mm and 3.0~mm is provided in the bottom panel.}
\label{fig:OpacityCompare}
\end{figure}

The FEOSAD code computes the evolution of two dust populations: small dust with a fixed dust size distribution between $5\times 10^{-3}~\mu$m and $1.0~\mu$m, and  grown dust with an adaptive dust size distribution between $1.0~\mu$m and $a_{\rm max}$. 
To calculate the spectral indices in the dust millimeter emission, we use the frequency dependent dust absorption opacities from \citet{Skliarevskii2021}. For the reader convenience, Figure~\ref{fig:OpacityCompare} presents these opacities as a function of the maximum dust size $a_{\rm max}$. For comparison, we also provide the corresponding opacities derived using the ``OpacityTool'' of \citet{Woitke2016} for the same dust composition. The bottom panel displays the comparison between our values and those from Woitke et al. at 1.3~mm and 3.0~mm, namely the wavelengths used to calculate the spectral index in this work. As can be seen, the agreement is excellent.

\section{Cooling and heating times}
\label{Append:coolheat}

\begin{figure}
\includegraphics[width=\linewidth]{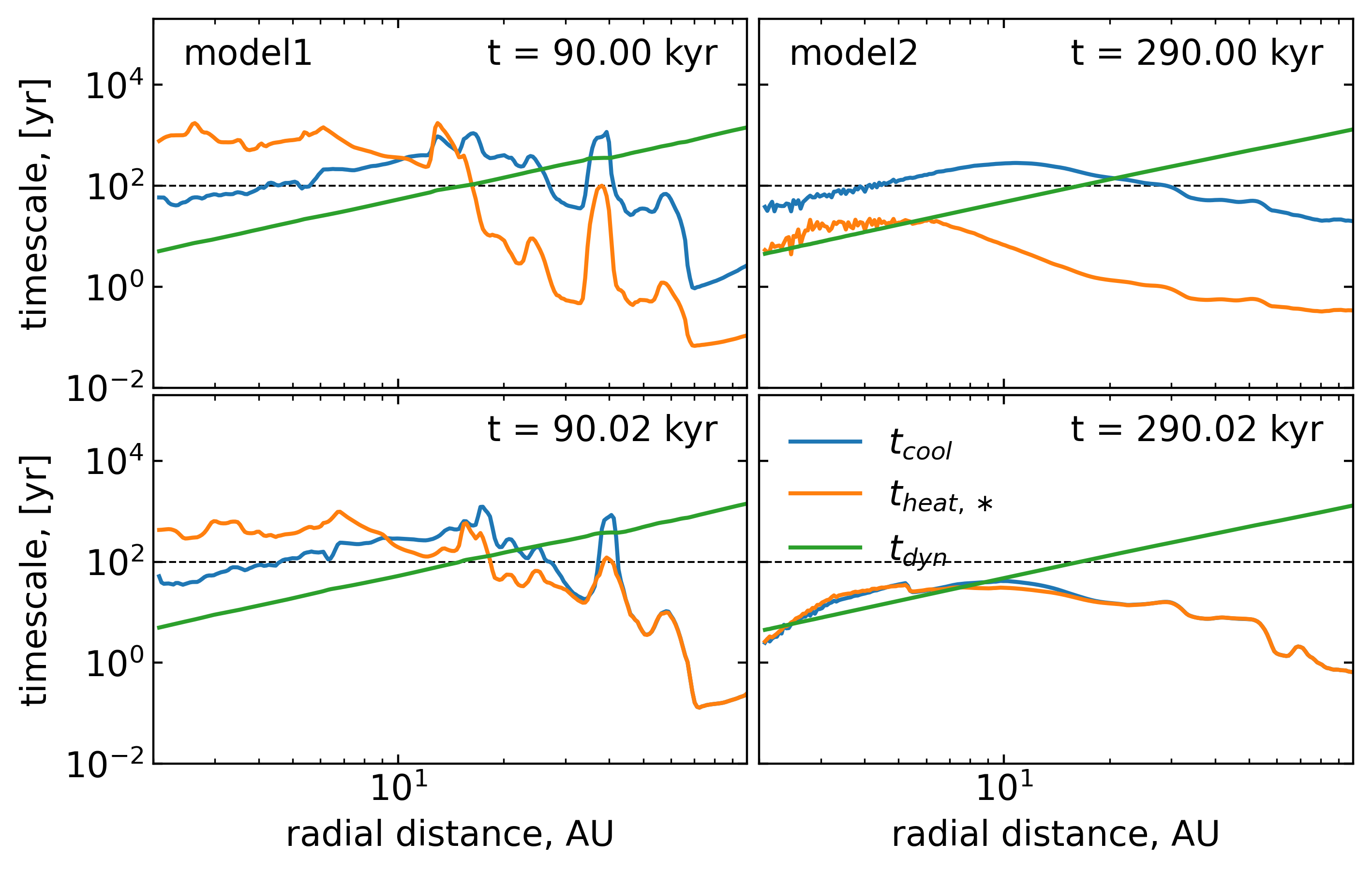}
\caption{Pertinent timescales in model~1 (left column) and model~2 (right column) at the onset of the burst (top row) and 20~yr after the onset of the burst (bottom row). The orange, blue, and green lines correspond to the stellar heating time, dust cooling time, and dynamical time. The horizontal dotted line indicates the burst duration. }
\label{fig:coolheat}
\end{figure}

Here we present the heating and cooling times in our burst models 1 and 2 and compare them with the dynamical timescale and the burst duration. The cooling time $t_{\rm cool}$ due to dust thermal emission and the heating time $t_{\rm heat,\ast}$ due to stellar irradiation are calculated as
\begin{equation}
    t_{\rm cool} = {e \over \Lambda},  \,\,\, t_{\rm heat,\ast} =  {e \over \Gamma},
\end{equation}
where the cooling and heating rates $\Lambda$ and $\Gamma$ are defined as
\begin{equation}
\Lambda=\frac{8\tau_{\rm P} \sigma T_{\rm mp}^4 }{1+2\tau_{\rm P} + 
{3 \over 2}\tau_{\rm R}\tau_{\rm P}}, \,\,\, 
\Gamma=\frac{8\tau_{\rm P} \sigma T_{\rm irr}^4 }{1+2\tau_{\rm P} + {3 \over 2}\tau_{\rm R}\tau_{\rm
P}},
\end{equation}
where $T_{\rm mp}$ is the disk temperature in the midplane (calculated from the vertically integrated pressure and gas surface density using the ideal equation of state) and $T_{\rm irr}$ is the irradiation temperature at the disk surface 
determined from the stellar and background black-body irradiation as
\begin{equation}
T_{\rm irr}^4=T_{\rm bg}^4+\frac{F_{\rm irr}(r)}{\sigma},
\label{fluxCS}
\end{equation}
where $\sigma$ is the Stefan-Boltzmann constant and  $F_{\rm
irr}(r)$ is the radiation flux (energy per unit time per unit surface
area)  absorbed by the disk surface at radial distance  $r$ from the
central star. The latter quantity is calculated as 
\begin{equation}
F_{\rm irr}(r)= \frac{L_\ast}{4\pi r^2} \cos{\gamma_{\rm irr}},
\label{fluxF}
\end{equation}
where $\gamma_{\rm{irr}}$ is the incidence angle of radiation arriving at the disk surface (with respect to the normal) at radial distance $r$ \citep[see for details][]{2010VorobyovBasu}.
The background temperature is set equal to $T_{\rm bg}=15$~K. The stellar luminosity $L_{\ast}$ is the sum of the accretion and photospheric luminosities.

Figure~\ref{fig:coolheat} presents the radial profiles of the cooling and heating times in models~1 and 2, which are calculated immediately after the onset of the burst (top row) and twenty years through the burst (bottom row). We note that these data are not azimuthally averaged, but taken along a narrow radial cut with an azimuthal angle of $\phi=0^\circ$. Azimuthal averaging may wash out the features related to spiral arms, which we want to emphasize in the Figure.

Let us first consider model~2 with an evolved axisymmetric disk (right column). At the burst onset the heating time is shorter than that of cooling, implying that the disk starts to warm up. Twenty years after the burst onset the cooling and heating times are already similar, meaning that the disk has warmed up and reached a state where $T_{\rm mp} \approx T_{\rm irr}$. At the same time, $t_{\rm cool}$ and $t_{\rm heat,\ast}$ are shorter than the burst duration and are comparable to or shorter than the dynamical time. This means that the burst has a major impact on the disk in model~2. 

Let us now turn to a gravitationally unstable and non-axisymmetric disk in model~1. At the onset of the burst the heating time is shorter than that of cooling only in part of the disk beyond $\approx 15$~au. In the inner parts, $t_{\rm heat}>t_{\rm cool}$ implying that the burst has little effect on the disk. Cooling is balanced by viscous heating in these inner regions \citep{2018VorobyovAkimkin}. Twenty years after the burst onset, the cooling and heating times are shorter than both the burst duration and the dynamical time only beyond $\approx 25$~au, suggesting that the burst has an effect on the disk only in these outer regions. Furthermore, there is a local peak in $t_{\rm cool}$ and $t_{\rm heat}$ at $\approx 40$~au, which corresponds to a density enhancement in the spiral arm (see Figs.~\ref{fig:init} and \ref{fig:endburst}). The values of $t_{\rm cool}$ and $t_{\rm heat}$ in the vicinity of this peak are comparable to or higher than the burst duration. This explains why the water snowline has such a complex spatial morphology in model~1 by the end of the burst  (see Fig.~\ref{fig:endburst}). The burst is not long enough to warm up dense and optically thick spiral arms to the water sublimation temperature, whereas in the diluted inter-arm disk regions the temperature rises faster and water ice sublimates sooner. 

\section{Dust as a fluid}
\label{fluid}
{Whether or not the dynamics of dust grains can be described as a fluid depends on the mean free path of dust grains and on their number. In particular, for the fluid approach to be valid, the number of dust grains $N_{\rm d}$ must be sufficiently large to derive the mean  characteristics of dust flow (e.g., volume density, mean velocity) that are characterized by negligible statistical errors ($1/\sqrt{N_{\rm d}}$). The mean characteristics can be  calculated by integrating the six-dimensional  distribution function of dust grains over the velocity space. In addition, the mean free path of dust grains has to be much smaller than the size of a fluid element, so that losses or gains of dust particles via stochastic collisions between each other would not affect notably the mean dust characteristics. This condition also allows using an isotropic pressure in the case of gas fluids.  The fluid element itself must be much smaller than the characteristic size of the problem and it is natural to associate computational cells with these fluid elements in the grid-based codes, such as the FEOSAD code.}

\begin{figure}
\includegraphics[width=1\columnwidth]{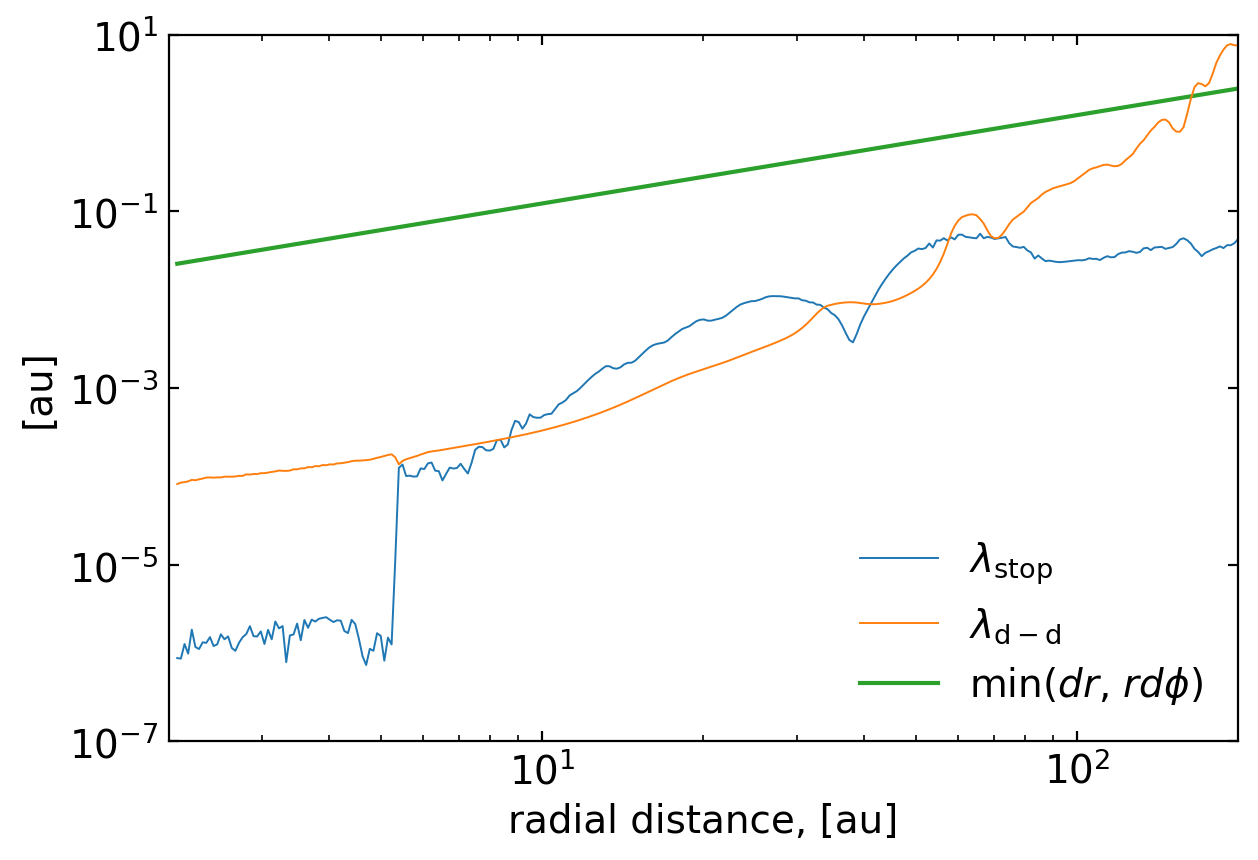}
\caption{Comparison of the mean free path of grain-to-grain collisions $\lambda_{\rm d-d}$ and dust stopping length $\lambda_{\rm stop}$ with the size of computational cells at a certain radial distance from the star. The data for model~2 at $t=290$~kyr were used to calculate these quantities.} 
\label{fig:dusthydro}
\end{figure}

{ Our model has two populations of dust grains: small and grown, both having a grain size distribution $dN_{\rm d} = C \cdot a^{-p} da$, where  $p = 3.5$ sets the steepness of the distribution and $C$ is a normalization factor. The latter can be determined based on the total mass of dust in each computational cell. It turns out that the number of grown dust particles $N_{\rm d}$ in each computational cell is on the order of $10^{25}$--$10^{33}$, depending on the maximum dust size $a_{\rm max}$, which is more than sufficient for constructing statistically accurate mean densities and velocities of dust grains.

In general, the mean free path of a particle in a medium composed of particles with similar sizes is defined as $\lambda =  1/(n \sigma)$,
where $n$ is the volume number density of particles and $\sigma$ is their collisional geometrical cross section. For a medium consisting of particles having a spectrum of sizes, as in our case of grain-to-grain collisions within each population, an average over the size distribution function has to be taken. The resulting mean free path for grown grain-to-grain collisions is written as
%, grain-to-gas, and gas-to-gas collisions are written as
\begin{equation}
    \lambda_{\rm d-d} =  \dfrac{1}{\langle n_{\rm d,gr}  \, \sigma_{\rm d,gr} \rangle},
\end{equation}
where $n_{\rm d, gr}$ is the volume number densities of grown dust in our model, derived using the information on the corresponding surface densities and vertical scale heights, and $\sigma_{\rm d,gr}$ is the geometrical cross-sections of grown dust. The brackets denote averaging that is taken over the entire population of grown dust grains from $a_\ast$ to $a_{\rm max}$ with a power index of $p=3.5$.

When considering the dust to gas collisions, it should be noted that the reduced mass that describes the momentum transfer between gas and dust due to individual collisions is negligible. In this case, it is appropriate to introduce the stopping length (instead of the mean free path) defined as
\begin{equation}
    \lambda_{\rm stop} = |\bl{u}_p - \bl{v}_p| \, t_{\rm stop}. 
\end{equation}
This quantity describes a distance that a dust grain can travel before its velocity becomes equal to that of gas. The values of gas and dust velocities, dust densities, and stopping time are taken from model~2 at $t=290$~kyr.

The resulting mean free path and stopping length are presented in Figure~\ref{fig:dusthydro} as a function of radial distance from the star. For comparison, we also plot the size of computational cells taken as a minimum of their radial and azimuthal extents. Clearly, the stopping length is one to several orders of magnitude smaller than the size of computational cells, meaning that the hydrodynamic description of dust flow is well justified (see Eq.~\ref{contDsmall}--\ref{momDlarge}). The grain-to-grain mean free path is generally several orders of magnitude smaller than the size of computational cells, but this trend diminishes at large distances where the dust surface density drops. This means that the use of the dust growth rate $\cal D$ in Eq.~(\ref{dustA}) is justified through most of the disk, except for its outermost regions where the disk mergers with a low-density surrounding medium. Since we are interested in processes occurring in the inner 100~au, the hydrodynamic approach is acceptable. We also note that dust growth diminishes in the envelope since $a_{\rm frag}$ becomes smaller than $a_\ast$ and the growth rate $\cal D$ becomes negligible there.  We  calculated the corresponding mean free paths for the small dust population and confirmed that the hydrodynamic approach is fulfilled to even a greater degree. 
}

\section{Dust to gas coupling}
\label{coupling}

\begin{figure}
\includegraphics[width=1\columnwidth]{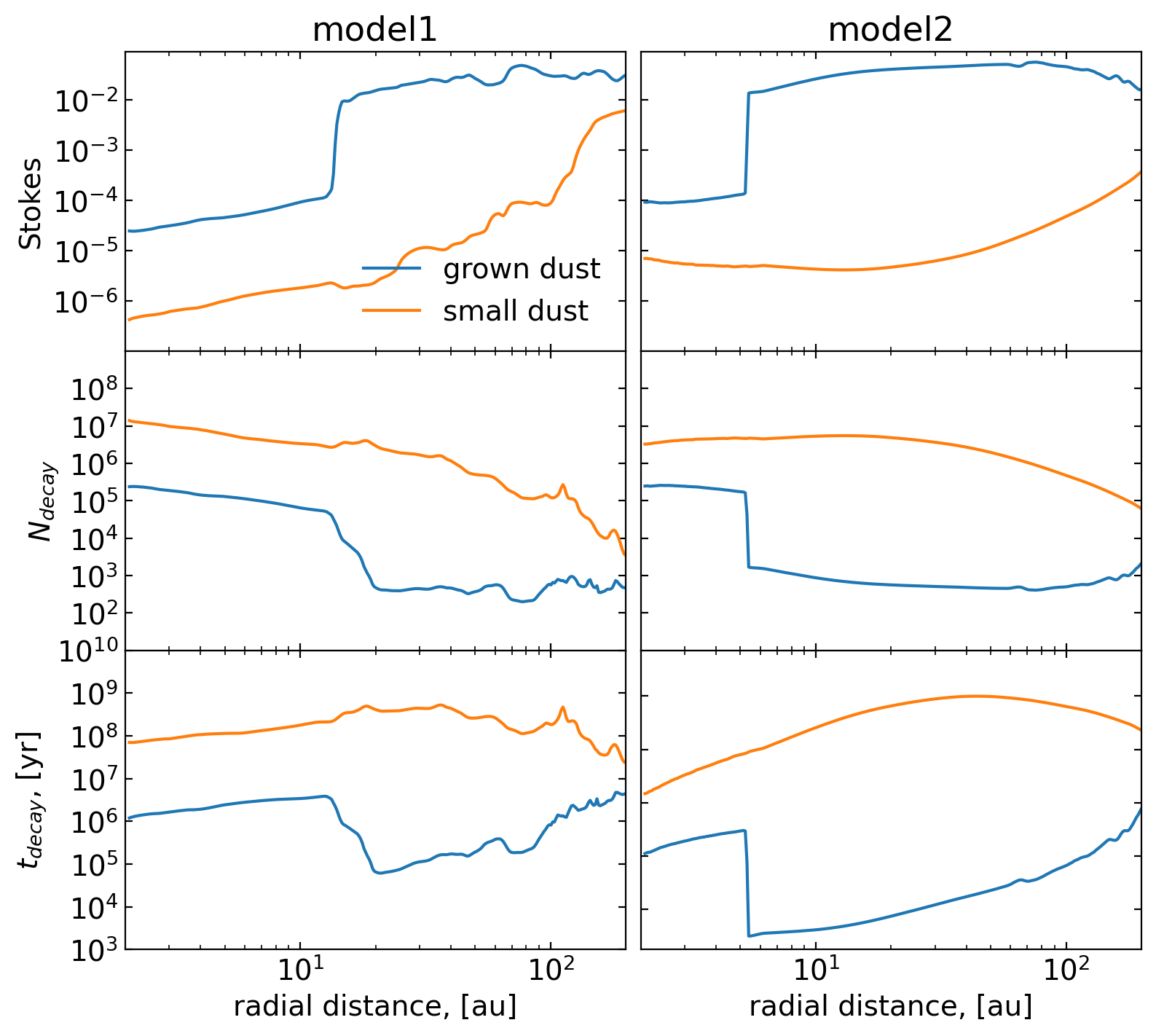}
\caption{Stokes number (top row), $N_{\rm decay}$ (middle row), and $t_{\rm decay}$ (bottom row) for the small and grown dust populations in model~1 (left column) and model~2 (right column). }
\label{fig:drift}
\end{figure}

{ In our model the dust grains are divided into two populations: small and grown, depending on their size. We further assume that small dust
is perfectly coupled to the gas while grown dust is partially decoupled (see Eqs.~\ref{contDsmall}--\ref{momDlarge}). Here, we show that this distinction is justified by calculating the time $t_{\rm decay}$ and the number of orbits $N_{\rm decay}$ for dust grains in each population to drift onto the star. The drift velocity of dust grains owing to gas pressure gradients in the disk can be written as follows \citep{Birnstiel2016}
\begin{equation}
u_{r,\mathrm{drift}} = - \dfrac{2 V_\mathrm{K} \, \mathrm{St} \, \eta_{\rm dev}}{1 + \mathrm{St}^{2}}, 
\end{equation}
where $V_{\rm K}$ is the Keplerian velocity and $\eta_{\rm dev}$ quantifies the deviation of the gas disk from the Keplerian pattern of rotation, the value of which we calculated based on the model's known pressure gradients { as
\begin{equation}
    \eta_{\rm dev} = - \dfrac{1}{2} \left( \dfrac{H_{\rm g}}{r} \right)^2 \dfrac{\partial \ln {\cal P} }{\partial \ln r}.
\end{equation}
We note that $\eta_{dev}$ is not to be confused with the desorption rates $\eta_{s}^{\rm sm}$ and $\eta_{s} ^{\rm gr}$ introduced in Eqs.~\eqref{eq:sig1}--\eqref{eq:sig3}. Since the radial distribution of $\eta_{\rm dev}$ turned out to be rather noisy owing to local variations in the gas pressure, we used mean values of $1.3\times 10^{-2}$ and $3.4\times 10^{-3}$ for models~1 and 2, respectively.}
The number of orbits $N_{\rm decay}$ can be calculated as
\begin{equation}
N_\mathrm{decay} = \dfrac{r}{u_{r,\mathrm{drift}} T_{\rm K}} = \dfrac{1 + \mathrm{St}^{2}}{4 \pi \mathrm{St} \, \eta},
\end{equation}
where $T_{\rm K}$ is the Keplerian orbital period. The time $t_{\rm decay}=N_{\rm decay} T_{\rm K}$ is calculated for the mass of the central star 0.36 and 0.44~$M_\odot$ at $t=90$ and $t=290$~kyr, respectively.

Figure~\ref{fig:drift} presents $\mathrm{St}$, $N_{\rm decay}$, and $t_{\rm decay}$ as a function of radial distance for models~1 and 2. Clearly, the small dust population can be considered strictly linked to gas, because its decay time is much longer than the typical lifetime of the disk (a few Myr). In this case, the continuity equation for small dust is sufficient to describe its evolution. On the contrary, the grown dust population may have a much shorter decay time, particularly beyond the snow line where the Stokes number increases. Hence, we need to solve for the momentum equation of grown dust in addition to the continuity equation.
}

\end{appendix}

\bibliographystyle{aa}
\bibliography{refs}
$$
$$
\end{document}